\documentclass[a4paper,aps,preprintnumbers,showpacs,twocolumn,superscriptaddress,nofootinbib,amsmath,amssymb]{revtex4-1}
\usepackage{bm,natbib,stmaryrd,times,hyperref,multirow,color}
\usepackage{ulem}
\usepackage{mathrsfs}
\hypersetup{
  colorlinks=true,        % false: boxed links; true: colored links
  linkcolor=blue,         % color of internal links
  citecolor=magenta,      % color of links to bibliography
  filecolor=magenta,      % color of file links
  urlcolor=red            % color of external links
}

\usepackage{graphicx}

\newcommand{\ie}{{i.e.,}~}
\newcommand{\eg}{{e.g.,}~}

\newcommand{\rem}[1]{}
\graphicspath{{figures_pdf/}}

\allowdisplaybreaks

\begin{document}

\title{New method for shadow calculations: Application to parameterised axisymmetric black holes}

\author{Ziri Younsi}
\affiliation{Institute for Theoretical Physics, Goethe-University,
  Max-von-Laue-Str. 1, 60438 Frankfurt, Germany}

\author{Alexander Zhidenko}
\affiliation{Centro de Matem\'atica, Computa\c{c}\~ao e Cogni\c{c}\~ao,
  Universidade Federal do ABC (UFABC), Rua Aboli\c{c}\~ao, CEP:
  09210-180, Santo Andr\'e, SP, Brazil}
\affiliation{Institute for Theoretical Physics, Goethe-University,
  Max-von-Laue-Str. 1, 60438 Frankfurt, Germany}

\author{Luciano Rezzolla}
\affiliation{Institute for Theoretical Physics, Goethe-University,
  Max-von-Laue-Str. 1, 60438 Frankfurt, Germany}
\affiliation{Frankfurt Institute for Advanced Studies, 
Ruth-Moufang-Str. 1, D-60438, Frankfurt am Main, Germany}

\author{Roman Konoplya}
\affiliation{Institute for Theoretical Physics, Goethe-University,
  Max-von-Laue-Str. 1, 60438 Frankfurt, Germany}

\author{Yosuke Mizuno}
\affiliation{Institute for Theoretical Physics, Goethe-University,
  Max-von-Laue-Str. 1, 60438 Frankfurt, Germany}

\begin{abstract}
Collaborative international efforts under the name of the Event Horizon
Telescope project, using sub- mm very long baseline interferometry, are
soon expected to provide the first images of the shadow cast by the
candidate supermassive black hole in our Galactic center, Sagittarius
A*. Observations of this shadow would provide direct evidence of the
existence of astrophysical black holes. Although it is expected that
astrophysical black holes are described by the axisymmetric Kerr
solution, there also exist many other black hole solutions, both in
general relativity and in other theories of gravity, which cannot
presently be ruled out. To this end, we present calculations of black
hole shadow images from various metric theories of gravity as described
by our recent work on a general parameterisation of axisymmetric black
holes [R. Konoplya, L. Rezzolla and A. Zhidenko, Phys. Rev. D
  \textbf{93}, 064015 (2016)]. An algorithm to perform general
ray-tracing calculations for any metric theory of gravity is first
outlined and then employed to demonstrate that even for extremal metric
deformation parameters of various black hole spacetimes, this
parameterisation is both robust and rapidly convergent to the correct
solution.
\end{abstract}
\maketitle

%--------------------------------------------------------------
\section{Introduction}
%--------------------------------------------------------------

It is now widely believed that at the centre of every galaxy resides a
supermassive black hole. Observational evidence, particularly for our own
Galactic black hole candidate, Sagittarius A* (Sgr A*), is
compelling \cite{Eckart:1996,Gillessen:2009} and supports the notion of
an object of enormous density, most likely a supermassive black hole,
residing in the innermost central region.

However, direct observation of an astrophysical black hole remains
illusive, and this is because of the existence of the event horizon, that
is, a surface limiting a region of spacetime beyond which neither matter
nor radiation can escape the gravity of the black hole. Outside this
surface, but still in close proximity to the event horizon, lies the
photon-capture region, where photons follow unstable orbits. Hence, when
observing a black hole directly, we expect to see a ``silhouette'' of
this photon region. 
%, since all emission originating from within it cannot reach an external
%observer.
Therefore, black holes are expected to be observed
as a ``shadow'' on the background sky \cite{Grenzebach:2014,
Grenzebach:2016,Cunningham:1973}.

It is anticipated that submillimeter very long baseline interferometry
(VLBI) observations of Sgr A* with the Event Horizon Telescope (EHT)
\cite{Huang:2007,Doeleman:2008,Goddi:2016} will soon yield the first
radio images of the shadow of the candidate black hole therein. The Black
Hole Camera project, in addition to other scientific activities,
participates actively to the investigation of the physics and
astrophysics of the black-hole candidate associated to Sgr A*. Particular
attention is dedicated to theoretical calculations of the shadows, whose
size and shape are sensitive to certain system parameters, in particular
the black hole mass and spin, as well as the orientation of the spin axis
of the black hole with respect to Earth (see, \eg
Ref. \cite{Falcke:2000}). Observations of this shadow would not only
provide very compelling evidence for the existence of an event horizon,
but also enable estimates to be placed on these system parameters.

Whilst astrophysical black holes are expected to be described by the Kerr
solution, there exist numerous black hole solutions in other theories of
gravity (see, for example, \cite{Stein:2014xba, Ayzenberg:2014aka} and
references therein). One cannot yet exclude the possibility of many of
the black hole metrics available in the literature and as such they are
all, in a sense, potential candidates. Rather than investigate all
possible theories of gravity and their corresponding black hole solutions
one at a time, it is expedient to instead consider a model-independent
framework within which any particular solution to any theory of gravity
may be parameterised through a finite number of modifiable parameters.
These parameters can then be chosen to measure deviations from the Kerr
metric and may be estimated from astrophysical
observations \cite{Vigeland:2011ji}. 

There is a simple reason why this avenue is a viable one, and although it
is quite obvious, it may be useful to recall it here. The problem of
defining the properties of the shadow does not require the choice of a
theory of gravity, but only of a well-behaved expression for the metric
tensor. This is because all that is ultimately needed to compute a shadow is
the solution of the geodesic equations. The latter obviously do not
require any assumption on the theory of gravity, but only a well defined
and regular definition of the metric tensor.

In Ref. \cite{Rezzolla:2014}, such a parametric framework was introduced
to describe the spacetime of spherically symmetric and slowly rotating
black holes in generic metric theories of gravity. The parameterisation
in \cite{Rezzolla:2014} is based on a continued fraction expansion in
terms of a compactified radial coordinate. Building upon the framework
of \cite{Rezzolla:2014} to also include axisymmetric spacetimes,
Ref. \citep{Konoplya:2016jvv} presented a parametric description of
axisymmetric black holes in generic metric theories of gravity. This new
parameterisation is based on a double expansion in both the radial and
polar directions of a general stationary and axisymmetric metric, and is
practically independent of any specific metric theory of gravity.
Although it was shown to accurately reproduce, with only a small number
of parameters, several different spacetime geometries, the question of
how many expansion orders in each direction are required to accurately
describe physical processes within this parametric framework was not
addressed in Ref. \citep{Konoplya:2016jvv}.

However, it is important to establish whether such a framework can
reproduce, to high precision, the strong field behaviour of geodesics in
the vicinity of the event horizon of different black hole
spacetimes. Calculating the black hole shadow through direct numerical
integration of the geodesic equations in the parameterised form of a
reference black hole metric (and repeating the calculation at successive
expansion orders), and subsequently comparing this with the shadow
obtained from the analytic form of the ``un-parameterised" metric,
provides a practical and stringent test of this framework. In addition,
ray-tracing calculations of the shadows cast by different black hole
solutions can provide insight into the practical performance of the
application of this parameterisation in astrophysical calculations
involving electromagnetic radiation.

Such a framework has several important applications:
\begin{enumerate}

\item To enable black hole solutions in many metric theories of 
gravity to actually be written in algebraic form and therefore
investigated using ray-tracing and radiative-transfer methods.

\item To represent all black hole solutions in terms of just a 
few parameters, distinguishing between solutions on this basis.

\item To constrain and potentially (physically) exclude 
black hole solutions from many theories of gravity with just a few key
observational parameters necessary to reproduce the shadow curve
(see \cite{Abdujabbarov:2015} for a general approach).

\end{enumerate}

In this first study we concern ourselves only with the shadow images
obtained from black holes in different metric theories of gravity. Since
the observed properties of radiation emanating from a black hole are
subject to the spacetime through which the radiation propagates, it is
prudent to first develop a method to ray-trace through a general
parameterised metric and investigate the accuracy of this
parameterisation.

Hence, we here numerically calculate the shadow boundary curve and
investigate, for several different spacetimes, the accuracy of the
parameterisation at various orders with respect to the original
un-parameterised form of the spacetime. Since the parameterisation
exactly reproduces Kerr in the equatorial plane, and in order to
adequately test the parameterisation, we consider near-extremal values of
all spacetime-specific deformation parameters. Different measures of the
accuracy of the expansion for each spacetime are presented and the
excellent convergence properties of the parameterisation are
demonstrated.

This paper is organised as follows. In Sec. \ref{sec:rtf} we describe the
ray-tracing formalism required to calculate geodesics within an arbitrary
metric parameterisation, where the expressions for such calculations are
derived explicitly. Section~ \ref{sec:pf} presents a short overview of
the axisymmetric parameterisation framework employed throughout this
paper. In Sec. \ref{sec:Kerr_Test} we apply this ray-tracing formalism to
several different known black hole solutions. Each parameterised
black hole solution is expanded to various orders and the resultant
black hole shadows are calculated and compared with the shadow from the
``exact" metric. Finally, Sec. \ref{sec:c} is devoted to the
conclusions.

%--------------------------------------------------------------
\section{Ray-Tracing Formalism}
\label{sec:rtf}
%--------------------------------------------------------------

In order to calculate the shadow image of a black hole, one must first
solve the geodesic equations in the background spacetime under
consideration. For the Kerr spacetime, there now exist several codes and
schemes to perform this task \cite[\eg][]{Fuerst:2004, Younsi:2012,
  Yang:2013, Chan:2013, Younsi:2015, Pu:2016, Dexter:2016}.

%In order to perform ray-tracing calculations for a series expansion of
%the metric coefficients, one can assume neither (i) conservation of
%energy, $E$, and angular momentum, $L_{\mathrm{z}}$ along the geodesic
%even though the metric is axisymmetric (given the expansion is finite and
%truncated) nor (ii) separability of the Hamilton-Jacobi equation, such
%that the geodesic equations may be solved as a quadrature problem given
%the conservation of both the Carter constant and the rest mass of the
%particle [$0$ for photons and $-1$ for particles in the $(-,+,+,+)$
%  convention].

As the order of the series expansion of the metric coefficients increases,
the expressions for these coefficients
grow rapidly in algebraic complexity.
Conventional methods to solve
the geodesic equations either through quadratures or by directly
integrating the geodesic equations are both impractical and inefficient,
as well as prone to large numerical errors.

Direct integration of the geodesic equations necessitates determining the
Christoffel symbols for the expanded metric at any given order. Given the
complexity of the expanded forms, this is impractical and the resultant
algebraic expressions can span hundreds of lines of code per Christoffel
symbol component. Moreover, such large expressions lead to a catastrophic
loss of numerical precision before the ray propagation even begins. A
naive approach would be to calculate the Christoffel symbols numerically,
but this again is inefficient since when evaluating partial derivatives
of the metric coefficients there are many repeated (as well as zero)
terms, and the computational overhead is significant. Since we seek to
minimise the number of operations needed to integrate the geodesic, we
must re-cast the geodesic equations in a form better-suited to satisfy
these requirements.

\subsection{Geodesic equations of motion}

For a given metric $g_{\alpha \beta}$, the Lagrangian may be written as:
\begin{equation}
2\mathscr{L} = g_{\alpha \beta}\,\dot{x}^{\alpha}\dot{x}^{\beta} \,,
\end{equation}
where an overdot denotes differentiation with respect to the affine
parameter, $\lambda$. Making $x^{\alpha}$ the variable of interest,
deriving the Euler-Lagrange equations and solving for $\ddot{x}^{\alpha}$ yields:
\begin{equation}
g_{\alpha\beta}\ddot{x}^{\beta}
= \partial_{\alpha}\mathscr{L}-\dot{g}_{\alpha \beta}\,\dot{x}^{\beta} \,,
\end{equation}
which may be re-written, upon raising and relabelling indices, as:
\begin{equation}
\ddot{x}^{\alpha} = g^{\alpha \beta}\left( \partial_{\beta}\mathscr{L} - 
\dot{g}_{\beta \gamma}\, \dot{x}^{\gamma} \right) \,, \label{geo_eqn}
\end{equation}
which are precisely the geodesic equations in a more succinct form.

In solving Eq.~(\ref{geo_eqn}) numerically, one must also employ the
result ${d} / {d}\lambda = \dot{x}^{\mu} \partial_{\mu}$
which enables Eq.~(\ref{geo_eqn}) to be written as:
\begin{equation}
\ddot{x}^{\alpha} = g^{\alpha \beta}\left( \partial_{\beta}\mathscr{L} -
\partial_{\mu} \ \! g_{\beta \gamma}\, \dot{x}^{\gamma} \dot{x}^{\mu}
\right) \,. 
\label{geodesic_equation}
\end{equation}

In general, one is only provided with the covariant metric components,
perhaps as a series expansion (as in this study), or on a grid of
simulation data (which would require interpolation between grid points,
\eg \cite{Meliani:2016}). Whilst one may determine the components of
the contravariant metric tensor from an algebraic expression in terms of
the determinant of the metric, for general metrics this is cumbersome,
computationally expensive and error-prone. Instead, we opt for numerical
lower-upper (LU) decomposition, being careful with singular regions such
as those near the event horizon, where the determinant can vanish.

\subsection{Application to axisymmetric spacetimes}

Although Eq.~(\ref{geodesic_equation})
represents the geodesic equations for any general metric tensor
$g_{\alpha \beta}$, in this study we restrict ourselves to metric
expansions of static and axisymmetric spacetimes expressed in Boyer-Lindquist coordinates, where the only off-diagonal metric
coefficient is $g_{t \phi}$.
As such, Eq.~(\ref{geodesic_equation}) may
be re-written in terms of the following system of second order ODEs:

\begin{eqnarray}
\overline{g} ~ \ddot{x}^{t} &=& g_{\phi\phi}\mathcal{T}_{t} - g_{t\phi}\mathcal{T}_{\phi} \,, \label{xdd_t} \\
g_{rr} ~ \ddot{x}^{r} &=& \partial_{r} \mathscr{L} - \partial_{\mu} g_{rr}~ \dot{x}^{r} \dot{x}^{\mu} \,, \label{xdd_r} \\
g_{\theta\theta} ~ \ddot{x}^{\theta} &=& \partial_{\theta} \mathscr{L} - \partial_{\mu} g_{\theta\theta}~ \dot{x}^{\theta} \dot{x}^{\mu} \,, \label{xdd_theta} \\
\overline{g} ~ \ddot{x}^{\phi} &=& g_{tt}\mathcal{T}_{\phi} - g_{t\phi}\mathcal{T}_{t} \,, \label{xdd_phi}
\end{eqnarray}
where
\begin{eqnarray}
\mathcal{T}_{t} &\equiv& \partial_{\mu} g_{tt} ~ \dot{x}^{t} \dot{x}^{\mu} \,, \\
\mathcal{T}_{\phi} &\equiv& \partial_{\mu} g_{\phi\phi} ~ \dot{x}^{\phi} \dot{x}^{\mu} \,, \label{T_phi}
\end{eqnarray}
and the index $\mu$ in Eqs. (\ref{xdd_t})--(\ref{T_phi}) ranges from $1$ to $2$ (\ie $r$, $\theta$) only.
Additionally, $g\equiv\mathrm{det}\left( g_{\alpha\beta}\right)=-g_{rr}g_{\theta\theta}(g_{t\phi}^{2}-g_{tt}g_{\phi\phi})$, where we have also defined
\begin{equation}
\overline{g} \equiv -g\left(g_{rr}g_{\theta\theta}\right)^{-1} 
= g_{t\phi}^{2}-g_{tt}g_{\phi\phi} \,.
\end{equation}

Solving Eqs. (\ref{xdd_t})--(\ref{xdd_phi}) directly, compared to solving
Eq.~(\ref{geodesic_equation}), has the advantage of both removing all vanishing
terms and expressing all equations in terms of covariant metric
components, thereby simplifying the resulting calculations\footnote{In
  particular, one may exploit the fact that, for axisymmetric spacetimes,
  the following identities hold: $g^{rr}g_{rr} =
  g^{\theta\theta}g_{\theta\theta}=1$, $ g^{tt} = -
  g_{\phi\phi} \, \overline{g}^{-1}$, $g^{t\phi} = g_{t\phi} \, \overline{g}^{-1}$, and
  $g^{\phi\phi} = - g_{tt} \, \overline{g}^{-1}$.}.
  
We note that the static and axisymmetric nature of the spacetime implies the conservation of energy, $E$, and angular momentum, $L_{\mathrm{z}}$, and consequently
Eqs. (\ref{xdd_t}) and (\ref{xdd_phi}) may be replaced by the following first-order ODEs:
\begin{eqnarray}
\overline{g} ~ \dot{x}^{t} &=& g_{\phi\phi}~E + g_{t\phi}~L_{\mathrm{z}} \,, \\
-\overline{g} ~ \dot{x}^{\phi} &=& g_{t\phi}~E + g_{tt}~L_{\mathrm{z}} \,,
\end{eqnarray}
where
\begin{eqnarray}
-E &= \dfrac{\partial \mathcal{L}}{\partial\dot{x}^{t}} =& g_{tt}\dot{x}^{t} + g_{t\phi}\dot{x}^{\phi} \,, \\
L_{\mathrm{z}} &= \dfrac{\partial \mathcal{L}}{\partial\dot{x}^{\phi}} =& g_{\phi\phi}\dot{x}^{\phi} + g_{t\phi}\dot{x}^{t} \,,
\end{eqnarray}
thereby reducing the number of ODEs to be integrated from $8$ to $6$.

Two additional constants of motion, namely the particle's rest mass, $\delta$ [equal to $0$ for photons and $-1$ for particles in the $(-,+,+,+)$ convention], and the Carter constant, $\mathcal{Q}$, enable the number of ODEs to be further reduced from $6$ to $4$ \cite{Carter:1968}.
However, Eqs. (\ref{xdd_r})--(\ref{xdd_theta}) are then replaced by first order ODEs which are of second degree in $\dot{x}^{r}$ and $\dot{x}^{\theta}$, respectively.
This introduces ambiguity in the signs of $\dot{x}^{r}$ and $\dot{x}^{\theta}$ at turning points in the geodesic motion due to the presence of square roots in the equations.
It is therefore more straightforward to simply integrate Eqs. (\ref{xdd_r})--(\ref{xdd_theta}) and avoid this issue altogether \cite{Pu:2016}. 

Although for any static and axisymmetric spacetime the energy and
angular momentum of a test point particle are conserved, for the purposes of comparing results at different expansion orders it
is more convenient not to enforce that $E$ and $L_{\mathrm{z}}$ must be conserved by construction.
For the same geodesic calculated at different orders of the expansion of the same spacetime, $E$, $L_{\mathrm{z}}$ and $\dot{x}^{t}$ will also be different at each expansion order.

Furthermore, when considering metrics written in more general (\eg Cartesian or modified and horizon-penetrating) coordinate systems, it is useful to numerically calculate all components of $x^{\alpha}$ and $\dot{x}^{\alpha}$ for practical general-relativistic radiative transfer calculations, \eg involving general-relativistic magnetohydrodynamical (GRMHD) simulation data \cite[e.g.][]{Meliani:2016}.
With $x^{\alpha}$ and $\dot{x}^{\alpha}$ fully computed, we calculate the values of $(E,L_{\mathrm{z}},\delta,\mathcal{Q})$ at every step of the geodesic integration. This enables us to check the accuracy of the integration by monitoring the conservation of these computed constants of motion with respect to their initial values at the beginning of the geodesic integration. For these reasons, in this study we numerically integrate Eqs. (\ref{xdd_t})--(\ref{xdd_phi}) directly.

Partial derivatives are evaluated using finite-difference representations
of the differential operators. The background spacetime is always
represented algebraically and in closed form in the parameterisation
scheme so in principle all metric coefficients may be evaluated to
machine precision. As such, we find that second-order centred finite
differencing with a step size between $10^{-4}\,M$ and $10^{-5}\,M$
(where $M$ is the black hole mass) is sufficient for the vast majority of
geodesic calculations considered in this paper. Occasionally, switching
to a fourth order method is necessary to maintain numerical precision in
problematic regions, \eg near the event horizon, polar regions or other
coordinate-dependent pathologies. In such regions, either forward or
backward finite-differencing methods are particularly useful.

Each geodesic is calculated to a precision of better than $10^{-9}\,M$
using a fourth order Runge-Kutta-Fehlberg integrator with adaptive step
sizing and fifth-order error control \citep{Press:1992}. If the input
spacetime were, for example, tabulated on a grid, then interpolation
between grid points would be required and thus higher order
finite-differencing methods would become necessary to preserve accuracy.

\subsection{Initial Conditions}

As is customary in ray-tracing calculations, an observer needs to be placed
at some distance from the source. In our calculations the observer is
positioned far from the black hole (\ie at $r_{\mathrm{obs}} =
10^{3}\,M$), where the spacetime is assumed to be essentially flat. The
observer's position is specified in Boyer-Lindquist (oblate spheroidal)
coordinates as $(r_{\mathrm{obs}}, \theta_{\mathrm{obs}},
\phi_{\mathrm{obs}})$.

The observer's image plane is a two-dimensional rectangular grid with
zero curvature, where each ray arrives perpendicular to the grid. The
initial conditions of each ray are then specified by transforming the
$(x,y)$ coordinates of the image plane into Boyer-Lindquist coordinates
in the black hole frame. The observer's $z$--direction is oriented along
the radial direction towards the black hole centre. After this
transformation, the coordinates of each pixel on the image plane are
expressed as follows:
\begin{eqnarray}
\ \! r^{2} &=& \sigma+\sqrt{\sigma^{2}+a^{2}Z^{2}} \,, \label{CartBLr} \\
\cos \theta &=& Z/r \,, \label{CartBLtheta} \\
\tan \phi &=& Y/X \,, \label{CartBLphi}
\end{eqnarray}
where
\begin{eqnarray}
X &\equiv& \mathcal{D}\cos\phi_{\mathrm{obs}}-x\sin\phi_{\mathrm{obs}} \,, \\
Y &\equiv& \mathcal{D}\sin\phi_{\mathrm{obs}}+x\cos\phi_{\mathrm{obs}} \,, \\
Z &\equiv& r_{\mathrm{obs}}\cos\theta_{\mathrm{obs}}+y\sin\theta_{\mathrm{obs}} \,,
\end{eqnarray}
and
\begin{eqnarray}
\sigma &\equiv& \left(X^{2}+Y^{2}+Z^{2}-a^{2}\right)/2 \,, \\
\mathcal{D}
&\equiv& \sin\theta_{\mathrm{obs}} \sqrt{r_{\mathrm{obs}}^{2}+a^{2}} -
y \cos\theta_{\mathrm{obs}}\,.
\end{eqnarray}

The components of the three-velocity of the ray are calculated through
differentiation of Eqs. (\ref{CartBLr})--(\ref{CartBLphi}), yielding:
\begin{eqnarray}
-\Sigma \ \dot{x}^{r} &=& r \mathcal{R} \sin\theta \sin\theta_{\mathrm{obs}} \cos \Phi+\mathcal{R}^{2} \cos\theta \cos\theta_{\mathrm{obs}} \,, \label{rdot_initial} \\
-\Sigma \ \dot{x}^{\theta} &=& \mathcal{R} \cos\theta \sin\theta_{\mathrm{obs}} \cos \Phi - r \sin\theta \cos \theta_{\mathrm{obs}} \,, \label{thetadot_initial}\\
\mathcal{R} \ \dot{x}^{\phi} &=& \sin\theta_{\mathrm{obs}} \sin \Phi \ \!  \mathrm{cosec}\theta \,, \label{phidot_initial}
\end{eqnarray}
where
\begin{eqnarray}
\Sigma &\equiv& r^{2} +a^{2}\cos^{2}\theta \,, \\
\mathcal{R} &\equiv& \sqrt{r^{2}+a^{2}} \,, \\
\Phi &\equiv& \phi-\phi_{\mathrm{obs}} \,.
\end{eqnarray}

Without loss of generality, the initial condition for the time coordinate
is set to be $t=0$ for all rays originating from the observer. The final
initial condition for $\dot{x}^{t}$ is calculated from the invariance of the
line element, yielding:
\begin{equation}
\dot{x}^{t} = \beta+ \sqrt{\beta^{2} + \gamma} \,,
\end{equation}
where
\begin{eqnarray}
\beta &\equiv& -\frac{g_{ti}\,\dot{x}^{i}}{g_{tt}} \,, \\
\gamma &\equiv& \frac{\delta - g_{ij}\,\dot{x}^{i}\dot{x}^{j}}{g_{tt}} \,.
\end{eqnarray}
Latin indices $\{i,j\}$ range from $1$ to $3$ (\ie $r$, $\theta$, $\phi$) and
denote the spatial components.

%%%%%%%%%%%% FIGURE 1 %%%%%%%%%%%%%
\begin{figure*}
\includegraphics[width=0.40\textwidth]{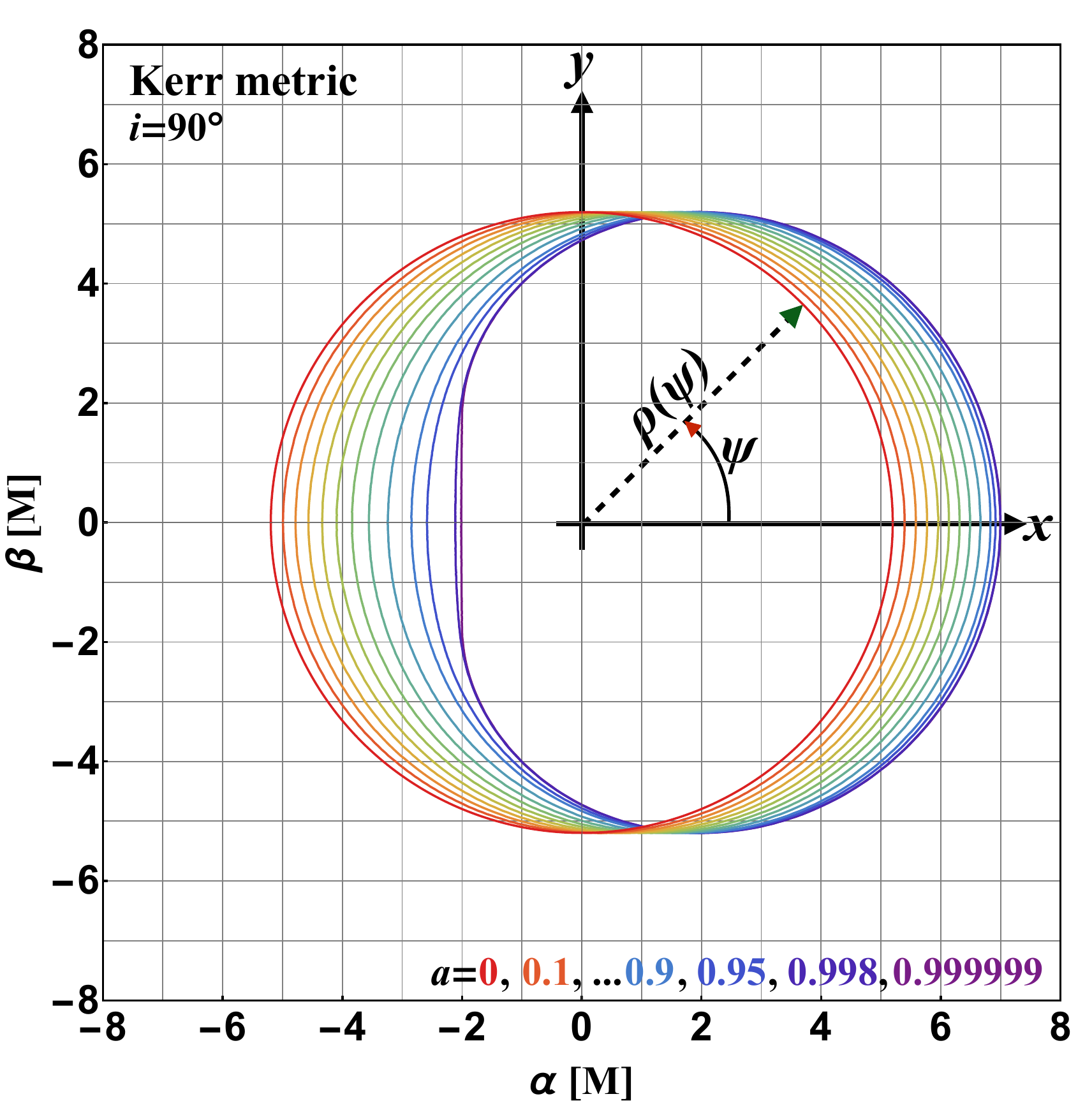} 
\hskip 1.75cm
\includegraphics[width=0.41\textwidth]{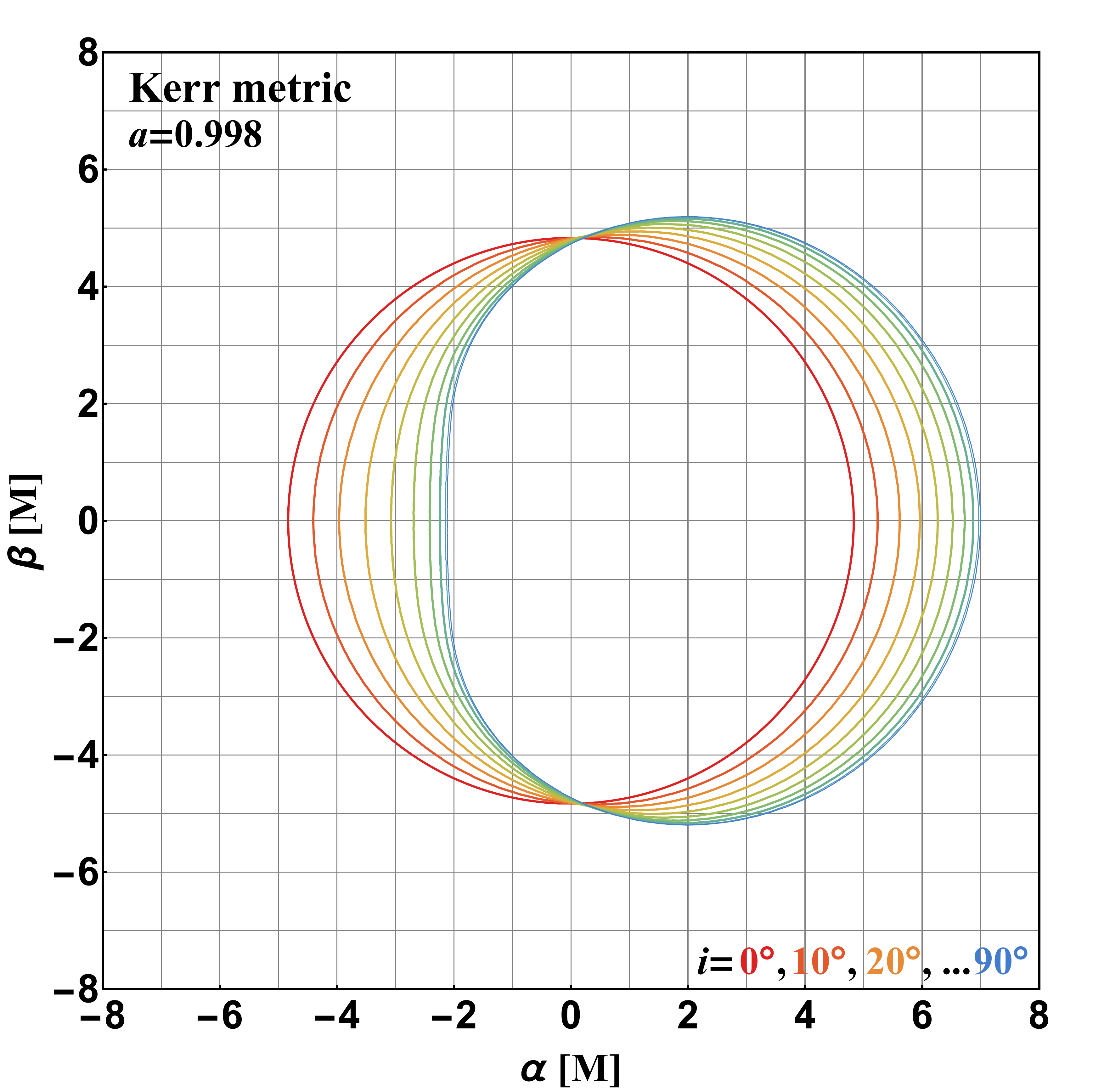}
\caption{Shadows cast by a Kerr black hole. \textit{Left panel:} As
  viewed by an observer at $i\equiv\theta_{\mathrm{obs}}=90^{\circ}$,
  with black hole spin parameters varied as $0$ (red, leftmost), $0.1$,
  $0.2$, \ldots, $0.9$, $0.95$, $0.998$ and $1-10^{-6}$ (purple,
  rightmost). \textit{Right panel:} Spin parameter fixed as $a=0.998$ and
  $i$ varied as $0^{\circ}$ (red, leftmost), $10^{\circ}$,
  $20^{\circ}$, \ldots, $90^{\circ}$ (blue, rightmost).}
\label{fig-1}
\end{figure*}
%%%%%%%%%% END OF FIGURE 1 %%%%%%%%%%

%--------------------------------------------------------------
\section{Parameterisation framework}
\label{sec:pf}
%--------------------------------------------------------------

We present here a brief overview of the parameterisation framework used
throughout this study. Further details and discussion may be found in
Ref. \citep{Konoplya:2016jvv}. We recall that in this parameterisation,
any axisymmetric black hole spacetime with mass $M$ and rotation
parameter $a$ can be represented by the following line element
\cite{Konoplya:2016jvv}
\begin{eqnarray}
\label{fixedmetric}
{d}s^2 &=&
-\dfrac{N^2-W^2\sin^2\theta}{K^2}{d}t^2-2Wr\sin^2\theta \ \! {d}t \ \! {d}\phi\\\nonumber&&
+K^2r^2\sin^2\theta \ \! {d}\phi^2
+S\left(\dfrac{B^2}{N^2}{d}r^2 +
r^2 {d}\theta^2\right)\,,
\end{eqnarray}
where
\begin{equation}\label{fixedmetriccond}
S\equiv\frac{\Sigma}{r^2} = 1+\frac{a^2}{r^2}\cos^2\theta\,, 
\end{equation}
and $N$,
$B$, $W$, $K$ are functions of the radial and polar (expanded in terms of
$\cos\theta$) coordinates as follows
\begin{subequations}\label{infiniteseries}
\begin{eqnarray}
B &=& 1+\sum\limits_{i = 0}^{\infty}B_i(r)(\cos\theta)^i\,,\\
W &=& \sum\limits_{i = 0}^{\infty}\dfrac{W_i(r)(\cos\theta)^i}{S}\,, \\
K^2&=& 1+\dfrac{aW}{r}+\frac{a^2}{r^2}+\sum\limits_{i = 1}^{\infty}\dfrac{K_i(r)(\cos\theta)^i}{S}\,,\\
N^2 &=& \left(1-\frac{r_0}{r}\right)A_0(r)+\sum\limits_{i = 1}^{\infty}A_i(r)(\cos\theta)^i\,,
\end{eqnarray}
\end{subequations}
where $r_0$ is the radius of the event horizon in the equatorial
plane\footnote{In Ref. \cite{Rezzolla:2014} the compactified radial
  coordinate $x \equiv 1 - r_0/r$ was introduced to simplify the
  expressions; while we could use such a coordinate here as well, we
  resort to the radial coordinate $r$ to ease the comparison with the
  original ``un-parameterised'' metrics.}.

We next expand the coefficients in terms of the radial coordinate as follows
\begin{subequations}
\begin{eqnarray}
B_i(r) &=& b_{i0}\frac{r_0}{r}+{\widetilde B}_i\frac{r_0^2}{r^2}\,,\\
W_i(r) &=& w_{i0}\frac{r_0^2}{r^2}+{\widetilde W}_i\frac{r_0^3}{r^3}\,,\\
K_{i>0}(r) &=& k_{i0}\frac{r_0^2}{r^2}+{\widetilde K}_i\frac{r_0^3}{r^3}\,,\\
A_0(r) &=& 1-\epsilon_0\frac{r_0}{r}+(a_{00}-\epsilon_0)\frac{r_0^2}{r^2}+\frac{a^2}{r^2}+{\widetilde A}_{0}\frac{r_0^3}{r^3}\,,\nonumber \\
A_{i>0}(r) &=& K_i(r)+\epsilon_{i}\frac{r_0^2}{r^2}+a_{i0}\frac{r_0^3}{r^3}+{\widetilde A}_i\frac{r_0^4}{r^4}\,,
\end{eqnarray}
\end{subequations}
where the tilded functions are given by 
\begin{subequations}
\begin{eqnarray}
{\widetilde B}_i \equiv \dfrac{b_{i1}}{1 + \dfrac{b_{i2}\left(1 -
    \frac{r_0}{r}\right)}{1 + \dfrac{b_{i3}\left(1 -
      \frac{r_0}{r}\right)}{1 + \ldots}}}\,,\\
{\widetilde W}_i \equiv \dfrac{w_{i1}}{1 + \dfrac{w_{i2}\left(1 -
    \frac{r_0}{r}\right)}{1 + \dfrac{w_{i3}\left(1 -
      \frac{r_0}{r}\right)}{1 + \ldots}}}\,,\\
{\widetilde K}_i \equiv \dfrac{k_{i1}}{1 + \dfrac{k_{i2}\left(1 -
    \frac{r_0}{r}\right)}{1 + \dfrac{k_{i3}\left(1 -
      \frac{r_0}{r}\right)}{1 + \ldots}}}\,,\\
{\widetilde A}_i \equiv \dfrac{a_{i1}}{1 + \dfrac{a_{i2}\left(1 -
    \frac{r_0}{r}\right)}{1 + \dfrac{a_{i3}\left(1 -
      \frac{r_0}{r}\right)}{1 + \ldots}}}\,.
\end{eqnarray}
\end{subequations}
Any approximation given by the above form of the metric is characterised
by two orders: the order of expansion in $\cos\theta$ ($m$) and the order
of the radial expansion ($n$). Specifying a finite number $m$ means
discarding all higher orders of the expansion, \ie we set $B_{i>m}=0$,
$W_{i>m}=0$, $K_{i>m}=0$ and $A_{i>m}=0$. As noted in
\cite{Konoplya:2016jvv}, it is not always possible to choose $a_{ij}=0$,
for any given $j>1$, in a consistent manner. The same applies to the
other continued-fraction coefficients, $b_{ij}$, $w_{ij}$, and
$k_{ij}$. This is why, in some cases, not all of the coefficients in the
radial expansion of order $n$ vanish for $j>n$, but their exact values
are substituted only for $j\leq n$.

%--------------------------------------------------------------
\section{Testing the parameterisation with shadow calculations}
\label{sec:Kerr_Test}
%--------------------------------------------------------------

Unless stated explicitly otherwise, hereafter the observer is positioned
at $i\equiv\theta_{\mathrm{obs}}=\pi/2$, \ie in the equatorial plane of a
rotating black hole. This provides the most extreme test of the effects
of gravitational lensing on the size and shape of the shadow image. In
calculating each shadow, due to the top-bottom symmetry of the image one
need only calculate the upper-half of the shadow and simply reflect this
in the observer's $x$-axis.

Consider the left panel of Fig.~\ref{fig-1}. A black hole shadow may be
represented as a closed parametric curve of radius $\rho(\psi)$, where
$\psi=[0,\pi]$, since the shadow is symmetric about the $x$-axis. The
interval $\psi$ is divided into $10^{3}$ equally-spaced points, and for
each value of $\psi$ bisection is performed along $\rho$ until convergence with
the shadow boundary is reached. 
Since the shadow is a single closed curve and $\psi$ is fixed, the bisection is one-dimensional and always around a single unknown but real and bounded point.

The bisection begins with the inner boundary placed at $(\alpha,\beta)=(0,0)$ and the outer boundary placed at the outermost $(\alpha,\beta)$ value for the particular value of $\psi$ chosen, \eg for Fig.~\ref{fig-1}, when choosing $\psi=45^{\circ}$ the inner and outer boundaries are placed at $(0,0)$ and $(8,8)$ respectively.
At both points a ray is fired towards the black hole and whether each ray is captured by the black hole (interior to the shadow) or escapes to infinity (exterior to the shadow) is determined.
Convergence is then defined as when the rays are within $10^{-6}~M$ of the shadow boundary, \ie bisection continues until the bisection step size is smaller than $10^{-6}~M$.

Before calculating shadows from more complicated expansions of metrics,
it is instructive to consider first the shadow from a Kerr black hole,
whose line element may be written in Boyer-Lindquist coordinates as:
\begin{eqnarray}
{d}s^{2} &=& -\left(1-\frac{2Mr}{\Sigma} \right){d}t^{2} +
\frac{\Sigma}{\Delta}{d}r^{2} + \Sigma \ \! {d}\theta^{2} \nonumber \\ &&
- \frac{4a M r \sin^{2}\theta}{\Sigma} {d}t \ \! {d}\phi +
\frac{\mathcal{A}\sin^{2}\theta}{\Sigma}{d}\phi^{2} \,,
\end{eqnarray}
where
\begin{eqnarray}
\Delta &\equiv& r^{2} - 2Mr + a^{2} \,, \\
\mathcal{A} &\equiv& \Sigma \ \! \Delta + 2 M r \left( r^{2} + a^{2} \right) \,.
\end{eqnarray}

Figure~\ref{fig-1} presents shadows cast by a Kerr black hole for a fixed
observer position and varying spin parameter (left panel) and for a fixed
extremal spin parameter and varying observer inclination angle (right
panel). It can be seen that increasing the black hole spin shifts the
shadow image to the right, rendering it more asymmetric and sharpening
the deviation from a circular shape, while the vertical extent of the
shadow remains unchanged (right panel). Fixing instead the spin parameter
to $a = 0.998$ and varying the observer inclination angle, both shifts
the shadow and increases its vertical extent (left panel). 

The parameterisation employed in this study exactly reproduces the Kerr
metric in the equatorial plane. Consequently, a stringent test of the
convergence properties of this parameterisation is best performed when
considering large values of the deformation parameters of the
parameterised metric. Whilst such large parameters may not be physically
realistic, they represent an important and practical ``stress test'' of
the parameterisation of each metric at different expansion orders in the
radial and polar coordinates.

%%%%%%%%%%%%%% FIGURE 2 %%%%%%%%%%%%%%%
\begin{figure*}
\raisebox{0.00cm}{\includegraphics[width=0.35\textwidth]{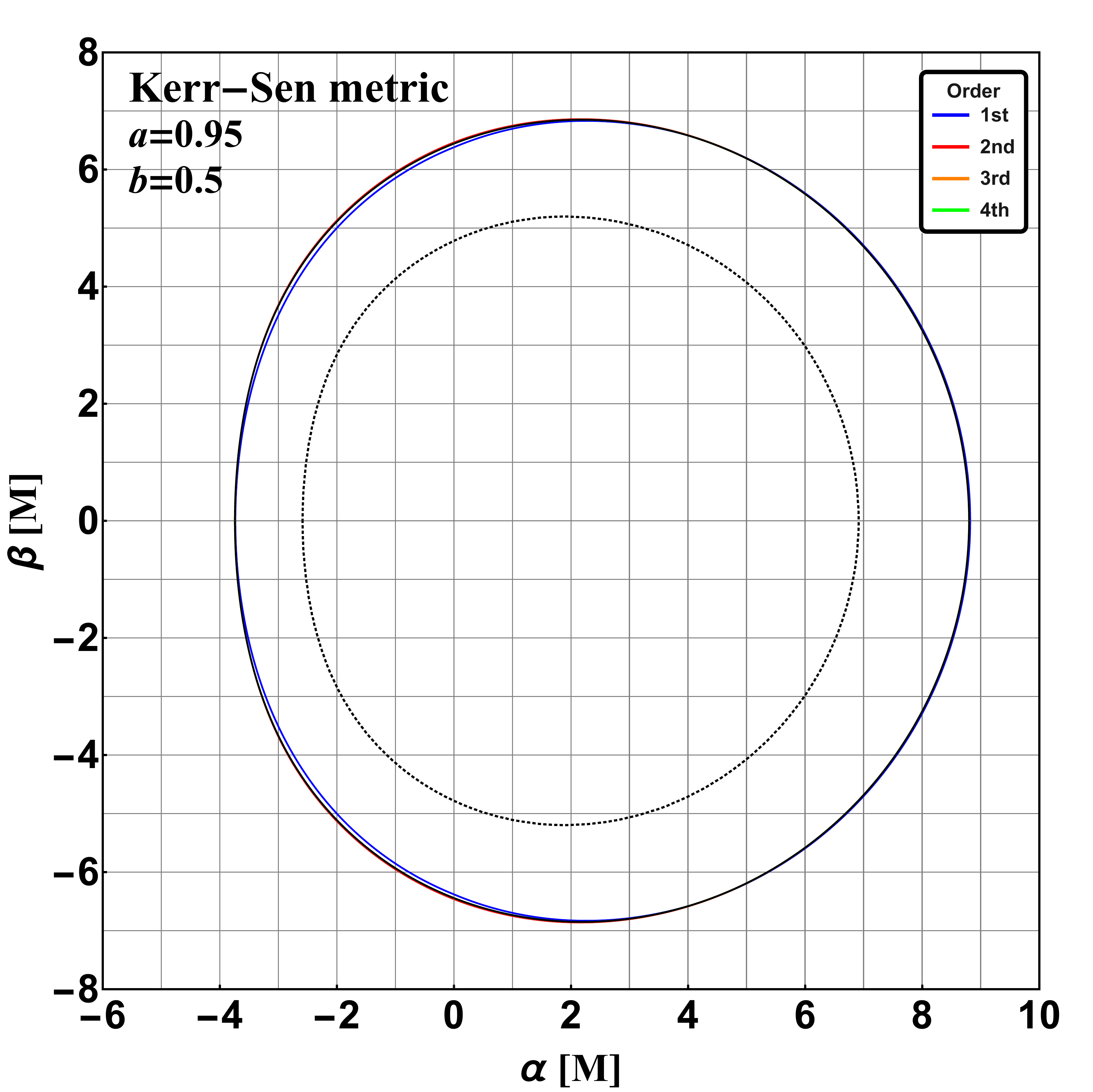}}
\hskip 1.0cm
\raisebox{0.75cm}{\includegraphics[width=0.4\textwidth]{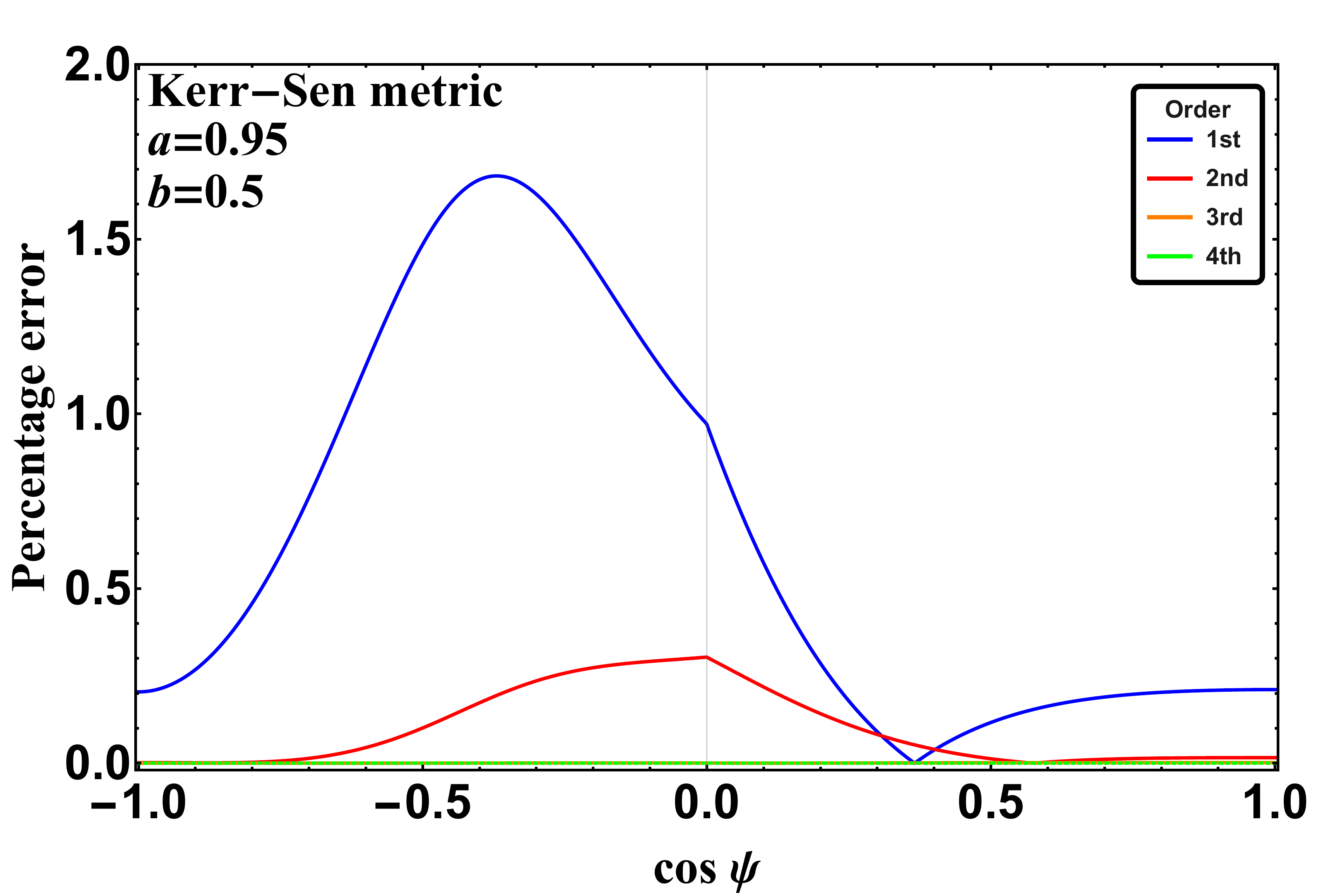}} 
\vskip 0.5cm
\raisebox{0.00cm}{\includegraphics[width=0.35\textwidth]{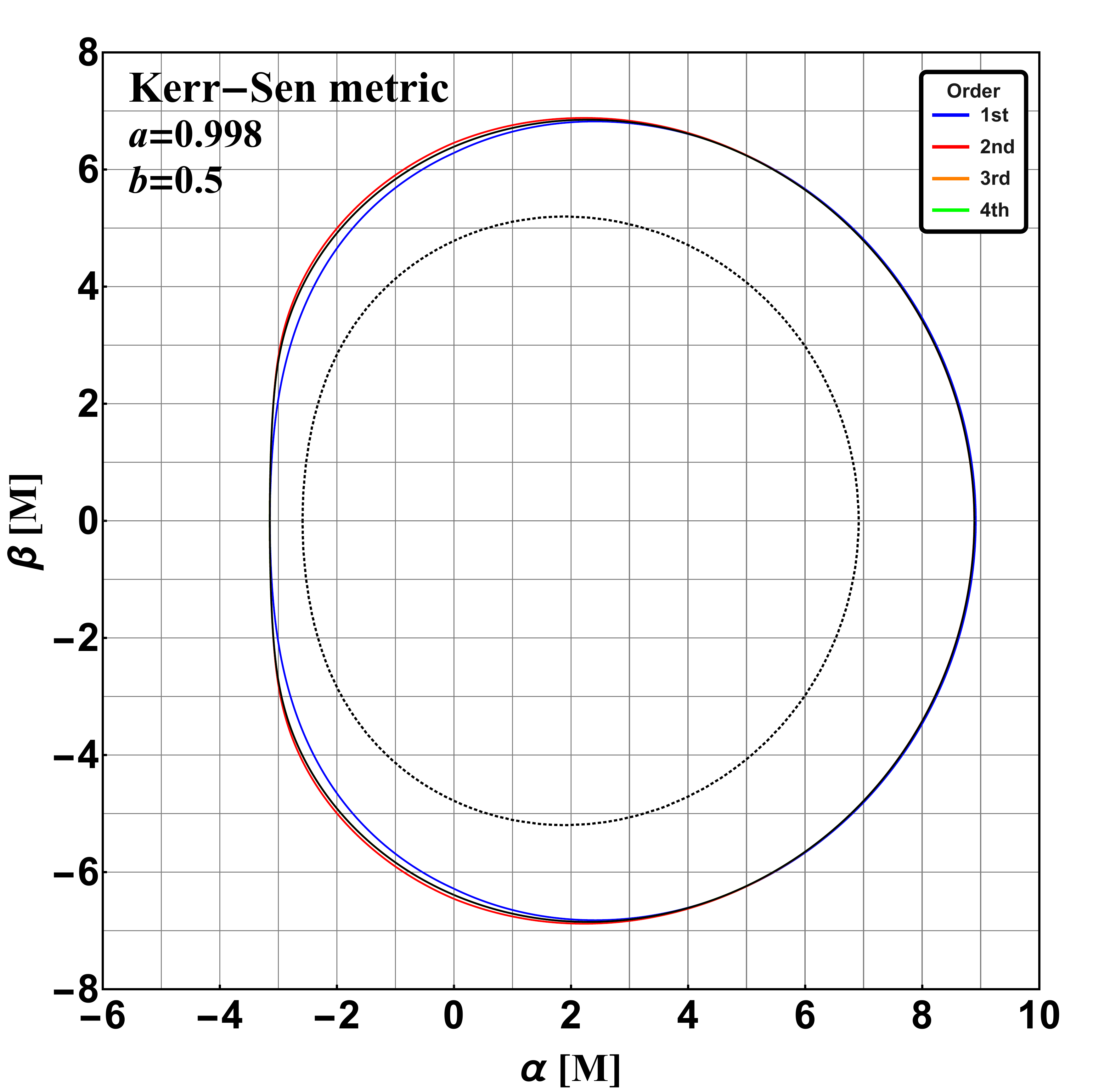}}
\hskip 1.0cm
\raisebox{0.75cm}{\includegraphics[width=0.4\textwidth]{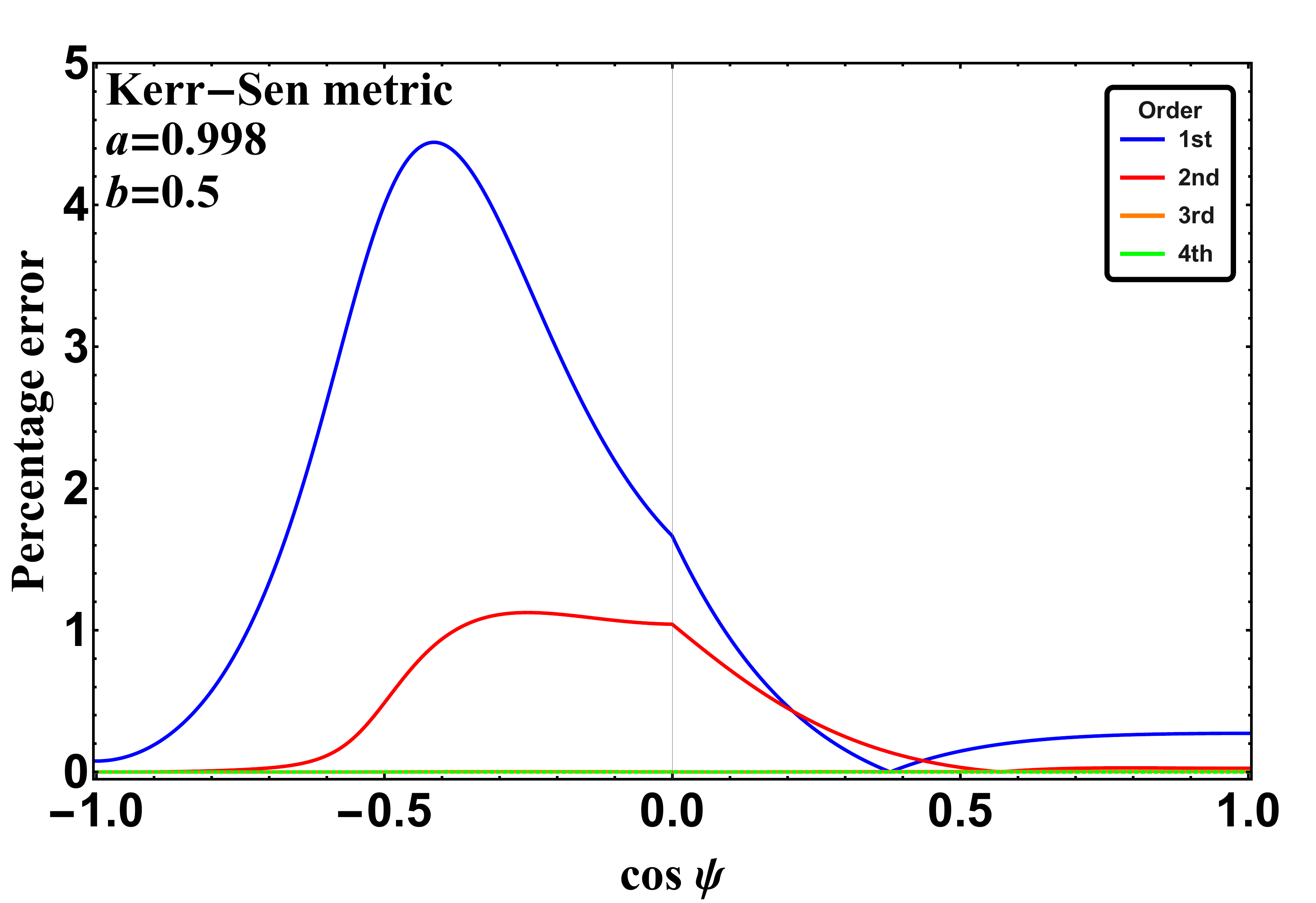}}
%% \includegraphics[width=0.385\textwidth]{Sen_a0p95_b0p5.pdf}
%% \includegraphics[width=0.597\textwidth]{Sen_a0p95_b0p5_error.pdf} \\
%% \vspace*{3mm}
%% \includegraphics[width=0.385\textwidth]{Sen_a0p998_b0p5.pdf}
%% \hspace{1.5mm}\includegraphics[width=0.587\textwidth]{Sen_a0p998_b0p5_error.pdf}
\caption{\textit{Left column:} Shadows from the \textit{radial} expansion
  of the Kerr-Sen metric for $b=0.5$ as viewed by an observer in the
  equatorial plane of the black hole for $a=0.95$ (\textit{top}) and
  $a=0.998$ (\textit{bottom}). The black curve represents the exact
  analytic form of the metric. Blue (first order), red (second order),
  orange (third order) and green (fourth order) curves represents the
  radial metric expansion. For comparison, the dotted curve shows the
  shadow from a Kerr black hole with the same spin parameter.
  \textit{Right column:} Corresponding percentage error plots of each
  expansion order with respect to the original Kerr-Sen metric as a
  function of $\cos\psi$ along the shadow boundary curve.}
\label{fig-2}
\end{figure*}
%%%%%%%%%%%% END OF FIGURE 2 %%%%%%%%%%%%

\subsection{General Testing Setup}

As discussed in \cite{Konoplya:2016jvv}, the axisymmetric expansion of
the metric may be in terms of $r$ (radial coordinate), $\cos\theta$
(where $\theta$ is the polar coordinate) or a combination of the two.  It
is important to remember that all axisymmetric black holes considered in
this study possess mirror symmetry, meaning that only even powers of
$\cos\theta$ are nonzero in the expansion. In this study, three distinct
black hole metrics are considered, each being represented as a series
expansion. The first is the Kerr-Sen metric \cite{Sen}, where an
exclusively radial expansion is employed. The second is the
Einstein-dilaton-Gauss-Bonnet (EDGB) \cite{Ayzenberg:2014aka} metric, 
which we expand in terms of $\cos\theta$, leaving the
coefficients in their exact form as functions of the radial coordinate.
The third metric is that proposed by Johannsen and Psaltis
\cite{Johannsen:2011dh}, wherein the expansion is performed in both the
radial coordinate and in $\cos\theta$.

In order to validate, both qualitatively and quantitatively, the
convergence properties and behaviour of the metric expansion, several
tests are performed. Naturally, given the original expression for a
metric and its series expansion at any particular order, one may visually
compare the shadow calculated from both forms of the metric, providing a
qualitative view of the performance of the expansion. This is the first
test, whereby the shadow calculated from the original metric is plotted
in black, and the shadows obtained from the expansion of this metric at
successive orders are over-plotted as coloured curves for comparison.

As it can be seen in Fig.~\ref{fig-1} (left panel), the shadow may be
represented as a closed parametric curve $\rho(\psi)$. In calculating the
shadow a bisection scheme is employed and the $\rho(\psi)$ value for each
sampled $\psi$ is recorded. The value of $\rho(\psi)$ is calculated both
for the original metric and for the various orders of expansion of this
metric. From this, the percentage error difference between the original
metric and any given order of the same expanded metric, \ie
$100\times|1-\rho(\psi)_{\rm analytic}/\rho(\psi)_{\rm expanded}|$ may be calculated as a function of
$\psi$ along the shadow boundary. Since $\psi=[0,\pi]$, the percentage
error is plotted as a function of $\cos\psi=[-1,1]$. This constitutes the
second test.

Since each shadow is a closed parametric curve, and given only the
upper-half of the shadow need be calculated, another measure of the
accuracy of the expansion is the area of the half-shadow. Since we store
the ($x$,\,$y$) coordinates of the shadow boundary curve we readily
obtain the area by calculating $\int_{x_{\mathrm{min}}}^{x_\mathrm{max}}
{d}x \, y(x)$ numerically. Finally, the third test of the expansion is
the percentage error difference between the half-shadow areas of the
original metric and its corresponding expansion.

\subsection{Kerr-Sen Metric}

An exact solution of the equations of motion corresponding to the
low-energy effective field approach of the heterotic string theory was
found by Sen in Ref. \cite{Sen}. This solution describes a charged,
axially-symmetric black hole (the Kerr-Sen black hole) \cite{Sen,Hioki}, whose
charge also introduces the presence of a scalar (dilaton) field $b$. The
Kerr-Sen (Sen) metric is a particular case of a more general
axion-dilaton black hole with a null Newman-Unti-Tamburino (NUT) charge
\cite{dilaton-BH}, and it can be described by the line element
(\ref{fixedmetric}) if one chooses the expansion in
\cite{Konoplya:2016jvv}, yielding
\begin{subequations}
\begin{eqnarray}
W &=& \frac{2a(\mu+b)(\sqrt{r^2+b^2}-b)}{r(r^2+a^2\cos^2\theta)}\,,\\
B^2 &=& \frac{r^2}{b^2+r^2}\,,\\
K^2 &=& \left(1+\frac{a^2\cos^2\theta}{r^2}\right)^{-1}
\left[\left(1+\frac{a^2}{r^2}\right)^2-\frac{a^2\sin^2\theta}{r^2}N^2\right]
\,, \nonumber \\ \\
N^2 &=& \frac{(\sqrt{b^2+r^2}-b)^2-2\mu(\sqrt{b^2+r^2}-b)+a^2}{r^2}\,,
\end{eqnarray}
\end{subequations}
where the ADM mass is now given by $M=\mu+b$. Hereafter we measure the
parameters $a$ and $b$ in units of $\mu$, \ie we choose $\mu=1$.

\begin{figure*}
\raisebox{0.00cm}{\includegraphics[width=0.35\textwidth]{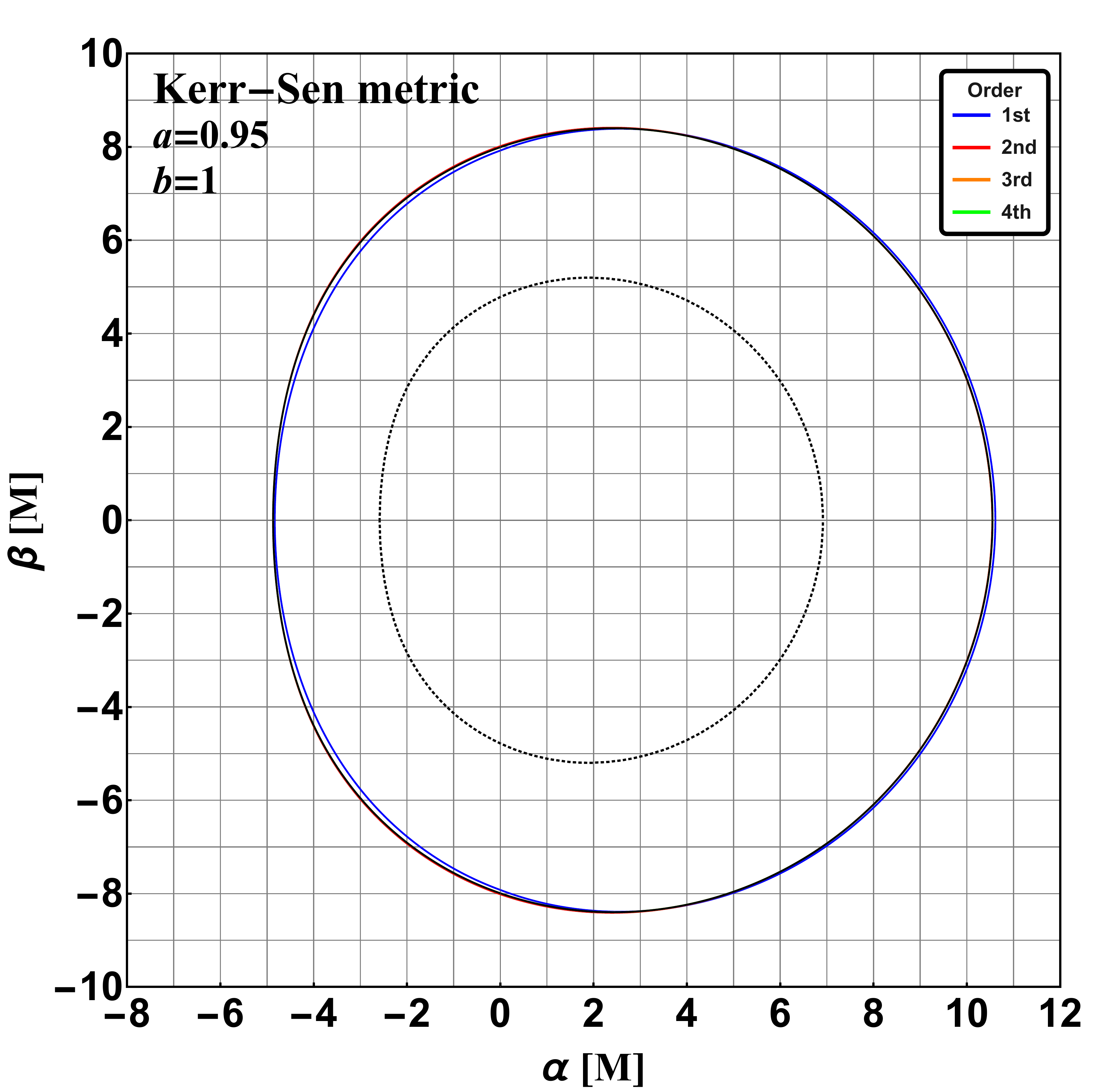}}
\hskip 1.0cm
\raisebox{0.75cm}{\includegraphics[width=0.40\textwidth]{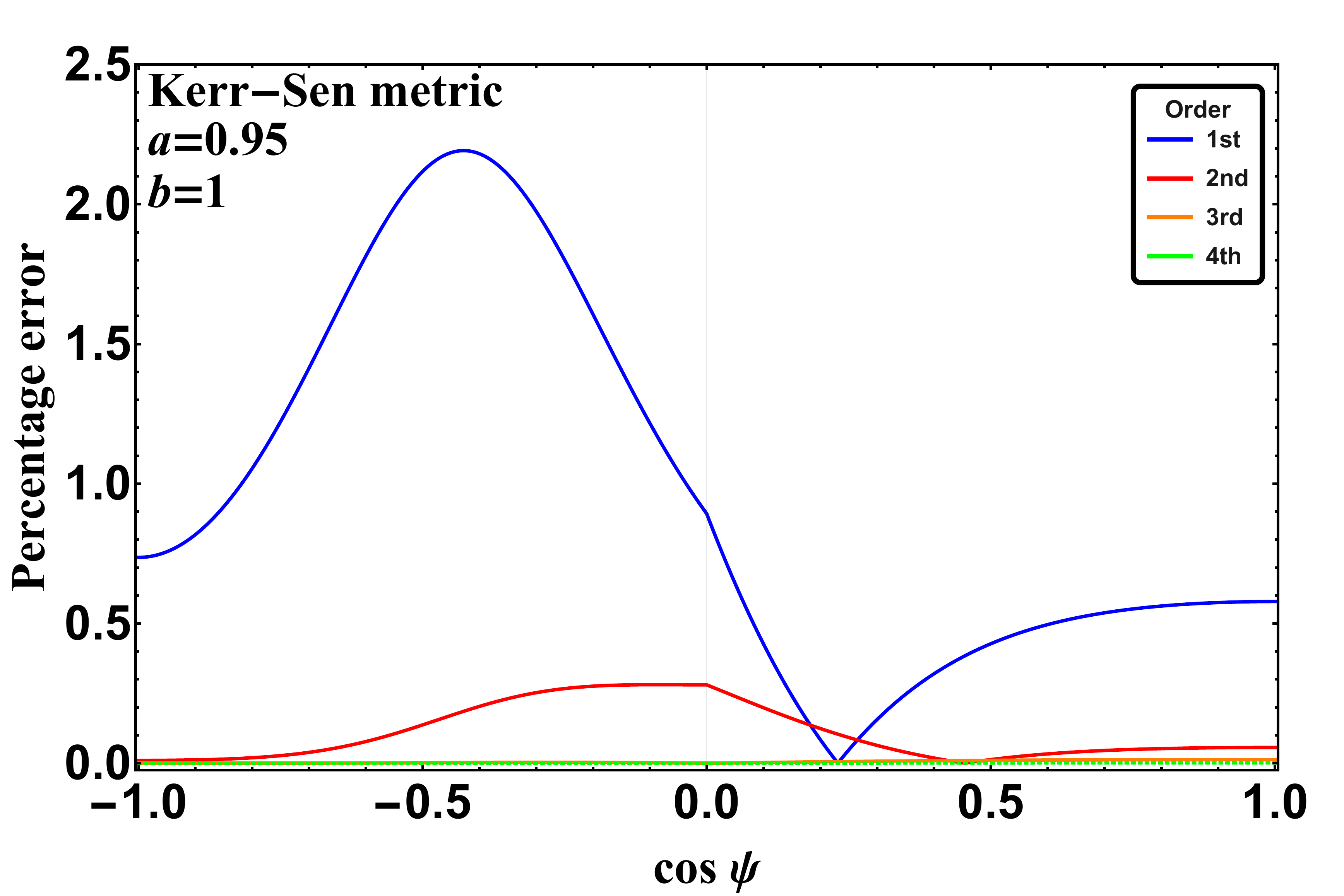}} 
\vskip 0.5cm
\raisebox{0.00cm}{\includegraphics[width=0.35\textwidth]{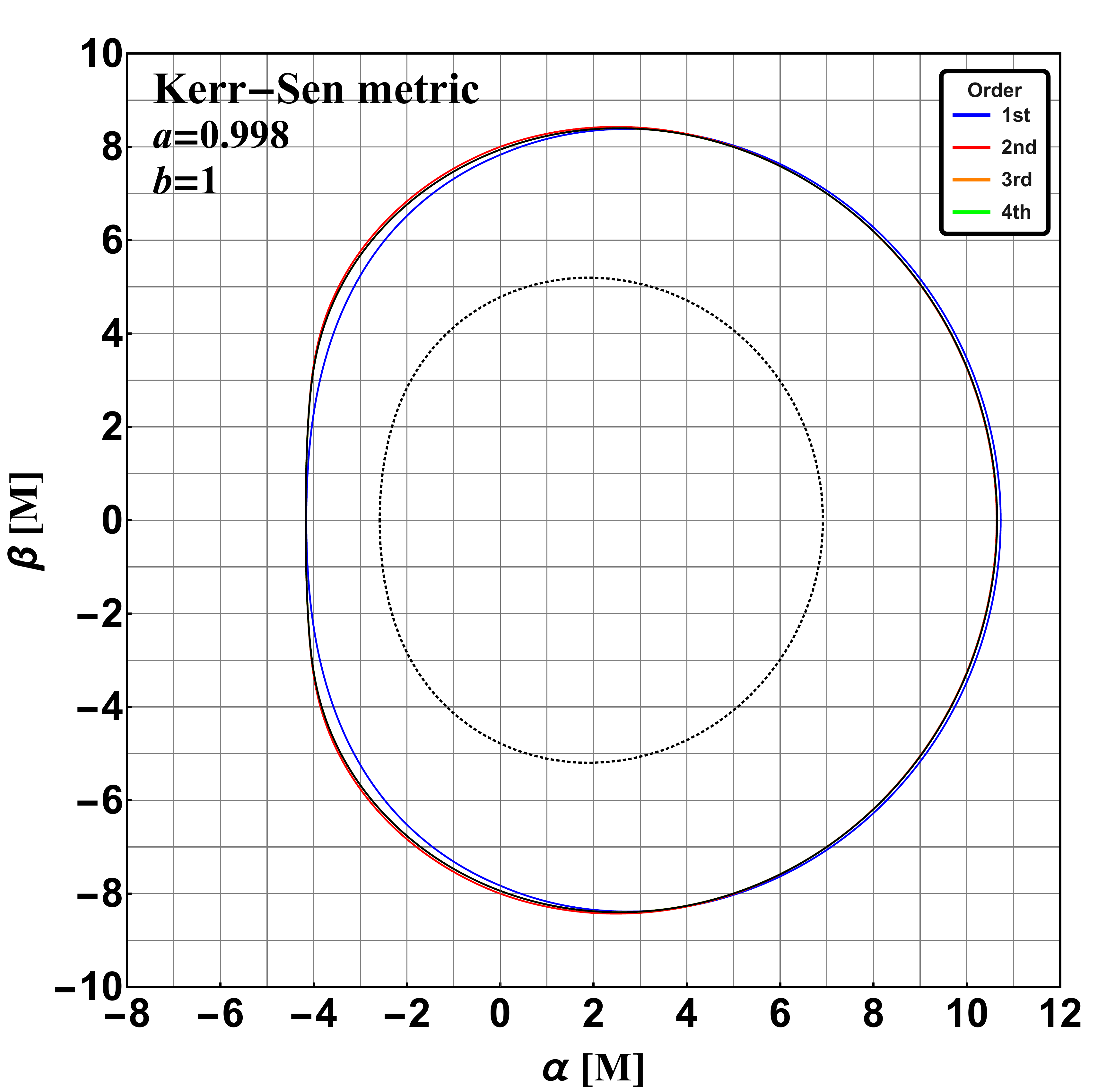}}
\hskip 1.0cm
\raisebox{0.75cm}{\includegraphics[width=0.40\textwidth]{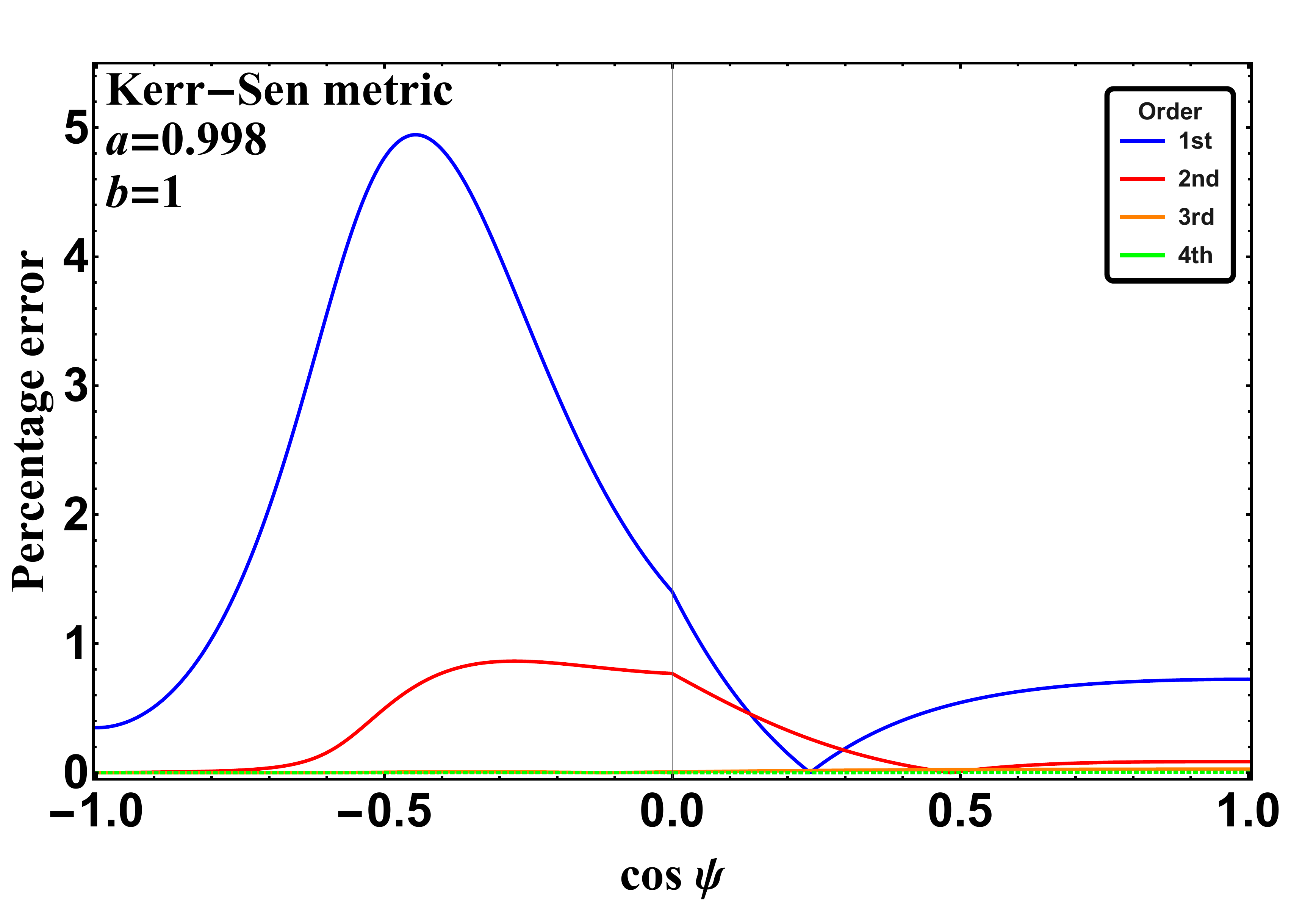}}
\caption{As in Fig.~\ref{fig-2}, but now the deformation parameter
  $b=1$.}
\label{fig-3}
\end{figure*}

The radial expansion of the Kerr-Sen metric is calculated from the first
through to the fourth order. The results of ray-tracing calculations of
the shadows from the radial expansion for the Kerr-Sen metric are
illustrated in the left panels of Figs.~\ref{fig-2} and \ref{fig-3} for
values of the dilaton field given by $b=0.5$ and $b=1$, respectively.

In the left panels of Fig.~\ref{fig-2}, the first-order expansion (blue
line) and the second order expansion (red line) are still visible and
most distinct from the exact shadow boundary (black line) towards the
left half of the image. For the right half of the image, on the other
hand, the agreement is excellent and improves as the shadow traverses the
equatorial plane ($y=0$ in the shadow image). The third and fourth-order
expansions (orange and green lines, respectively) cannot be seen and
overlay the exact black curve very well. As expected, the effect of
increasing the spin parameter from $0.9$ to $0.998$ is to further distort
the shadow image and slightly slow the convergence, as evidenced by the
blue and red curves being visually further apart from the black curve in
the $a=0.998$ case.

In the right panels of Fig.~\ref{fig-2} we show instead the relative
error (as a percentage) of each expansion order of the Kerr-Sen metric
relative to the exact metric; the various curves are plotted as a
function of $\psi$ along the shadow boundary. For $a=0.95$, the maximum
error in the first-order expansion (blue line) is $\sim 1.7~\%$. At
second order (blue line) this drops to $\sim 0.25~\%$ and by the third
(orange line) and fourth order (green line) the error is negligible and
thus the orange and green curves appear as horizontal lines.

As an additional ``stress-test'' of the parameterisation approach, we
consider in Fig.~\ref{fig-3} the more extreme deformations which follow
when considering a dilaton field with $b=1$. We note that the effect of
increasing $b$ is that of increasing the absolute size of the shadow (the
mass is proportional to $b$), so that the error inherent to each
expansion order also increases. Overall, the behaviour shown in
Fig. \ref{fig-3} is very similar to that of Fig.~\ref{fig-2}, with the
second and higher-order radial expansions again exhibiting excellent
convergence properties. Furthermore, and as found for lower values of
$b$, the error plots demonstrate that already at the second order the
error is everywhere below $0.3~\%$ and that the error at third and
fourth-order is very close to zero across the entire shadow.

As a final check of the accuracy and convergence of the series expansion,
the half-shadow area (\ie for $\psi \in [0,\pi/2]$) for both the exact
Kerr-Sen metric and its four different expansion orders is
calculated. These results are presented in Table~\ref{table:Sen}, where
four different values of the black hole spin parameter (\ie $a = 0.2,\,
0.5,\, 0.95,\, 0.998$) are considered both for $b=0.5$ and for $b=1$. The
table also reports as $\epsilon_{n,m}$ the relative error between the
area computed from the exact metric shadow and that obtained from the
same expanded metric shadow at order $n$ in the radial direction and $m$
in the polar direction.

In the case of slowly rotating black holes, \ie $a=0.2$, the error in the
area for the first-order expansion is $\sim 10^{-4}~\%$ and by the
fourth-order expansion it is within the precision used to calculate the
shadow and thus effectively zero. This trend also holds for $a=0.5$
(moderate spin), which is why shadow and error plots for these values
were not presented. For higher spins, the error at first order is larger
but still at the one percent level. More importantly, it can be seen that
the values of the area of the shadow rapidly converges to the ``true''
shadow area as the order of the expansion is increased.

Before concluding this Section dedicated to Kerr-Sen black holes we note
that the values chosen here both for the spin and for the dilaton field
are extreme and likely to be much larger than what would be found for an
astrophysical black hole. That being said, the convergence of the radial
expansion of the Kerr-Sen metric is both fast and highly accurate even in
these extremal cases.

%%%%%%%%%%%%%% FIGURE 3 %%%%%%%%%%%%%%%
%%%%%%%%%%%% END OF FIGURE 3 %%%%%%%%%%%%

\subsection{Einstein Dilaton Gauss-Bonnet Metric}

In $D>4$ spacetimes, where $D$ is the number of spacetime dimensions, the
second-order term in curvature, \ie the Gauss-Bonnet term, is the
dominant one. However, in four-dimensional spacetimes, like those
considered in this paper, the Gauss-Bonnet term alone is invariant and
leads to solutions of the Einstein equations which are not affected
unless the scalar field (dilaton) is coupled to the system. An
approximate metric for a rotating black hole in this system with such a
dilaton coupling, \ie in EDGB theory, was deduced in the regime of slow
rotation \cite{Ayzenberg:2014aka}. The metric was obtained in the form of
an expansion in terms of two small parameters: $\zeta$ and $\chi\equiv
a/M$. For finite values of these parameters, the metric has a divergence
at the Schwarzschild horizon $r=2M$. However, the equivalent form
proposed in Ref. \cite{Konoplya:2016jvv} removes this divergence and the
line element then reads
\begin{eqnarray}
\label{EDGB_metric}
{d}s^{2} &=& -\frac{f-w^{2}\sin^{2}\vartheta}{\kappa^2}{d}t^{2} +
\frac{\beta^2\sigma}{f}{d}\rho^{2} + \rho^{2} \sigma \,
     {d}\vartheta^{2} \nonumber \\ && - 2w \, \rho \sin^{2}\vartheta
     \, {d}t \, {d} \, \phi + \rho^{2} \kappa^2 \,
     \sin^{2}\vartheta \, {d}\phi^{2} \,, 
\end{eqnarray}
where the various terms in the metric are given by
\begin{subequations}\label{EDGB}
\begin{equation}
\begin{aligned}
\hspace{-2mm} \kappa^2 = 1 &+ \chi^{2}\frac{M^2}{\rho^2}\left[ 1 + \frac{2M}{\rho}\left( 1-\cos^2\vartheta \right) \right] \\
&+ \zeta\chi^{2}\left(\cos^2\vartheta - 1/3 \right) \frac{M^3}{\rho^3}\sum_{k=0}^{7}c_{k}\frac{M^k}{\rho^k} \! ,
\end{aligned}
\end{equation}
\begin{equation}
\begin{aligned}
\sigma = 1 &+ \chi^{2}\frac{M^2}{\rho^2}\cos^2\vartheta \\ 
&+ \zeta\chi^{2}\left(\cos^2\vartheta - 1/3 \right) \frac{M^3}{\rho^3} \sum_{k=0}^{7}c_{k}\frac{M^k}{\rho^k} \,,
\end{aligned}
\end{equation}
\begin{eqnarray}
\hspace{-18mm} w &=& 2\chi \frac{M^2}{\rho^2} + \frac{1}{15}\zeta\chi \frac{M^4}{\rho^4}\sum_{k=0}^{4}w_{k}\frac{M^k}{\rho^k} \,,
\end{eqnarray}
\begin{eqnarray}
f &=& 1-\frac{2M}{\rho}+\chi^{2}\frac{M^2}{\rho^2}+\zeta \frac{M^3}{6\rho^3}\left( 2-\chi^{2} \right) \\
&& \ \ \ + \ \zeta \frac{M^4}{\rho^4} \sum_{k=0}^{7}\left( \chi^{2}\cos^2\vartheta f_{k,1} + \chi^{2} f_{k,2} + f_{k,3} \right)\frac{M^k}{\rho^k} \,, \nonumber
\end{eqnarray}
\begin{eqnarray}
\beta^2 &=& 1 + \frac{M^2}{6\rho^2}\left( \chi^{2}-2 \right) \left( 3+\frac{8M}{\rho} \right) \\
&& \ \ \ + \ \zeta \frac{M^4}{\rho^4} \sum_{k=0}^{6}\left( \chi^{2}\cos^2\vartheta\beta_{k,1} + \chi^{2} \beta_{k,2} + \beta_{k,3} \right)\frac{M^k}{\rho^k}\,. \nonumber
\end{eqnarray}
\end{subequations}
The numerical values for the coefficients in the series in
Eqs. (\ref{EDGB}) may be found in Table~\ref{table:coefficients}.

In order to transform this metric into the form of
eq.~(\ref{fixedmetric}) we proceed as in Ref. \cite{Konoplya:2016jvv}.
We rewrite eq.~(\ref{EDGB_metric}) in the form (\ref{fixedmetric}) by
imposing that in terms of the new coordinates, $r$ and $\theta$, relation
(\ref{fixedmetriccond}) and the additional condition
$$\left(K^2-1-\dfrac{aW}{r}-\frac{a^2}{r^2}\right)\Biggr|_{\theta=\frac{\pi}{2}}=0
\,,$$ are fulfilled. Finally, we obtain the functions $B$, $W$, $K^2$,
and $N^2$ as an infinite series in terms of $\cos\theta$, as in
(\ref{infiniteseries}).  The lowest order is then given by expressions
(80) in Ref. \cite{Konoplya:2016jvv}, which we do not report here for
compactness. To provide a test of the polar expansion, the non-divergent
EDGB metric is expanded up to eighth order in powers of
$\cos\theta$. This is then compared with the exact form of the
non-divergent EDGB metric given in Eq.~(\ref{EDGB_metric}).

\begin{figure*}
\includegraphics[width=0.40\textwidth]{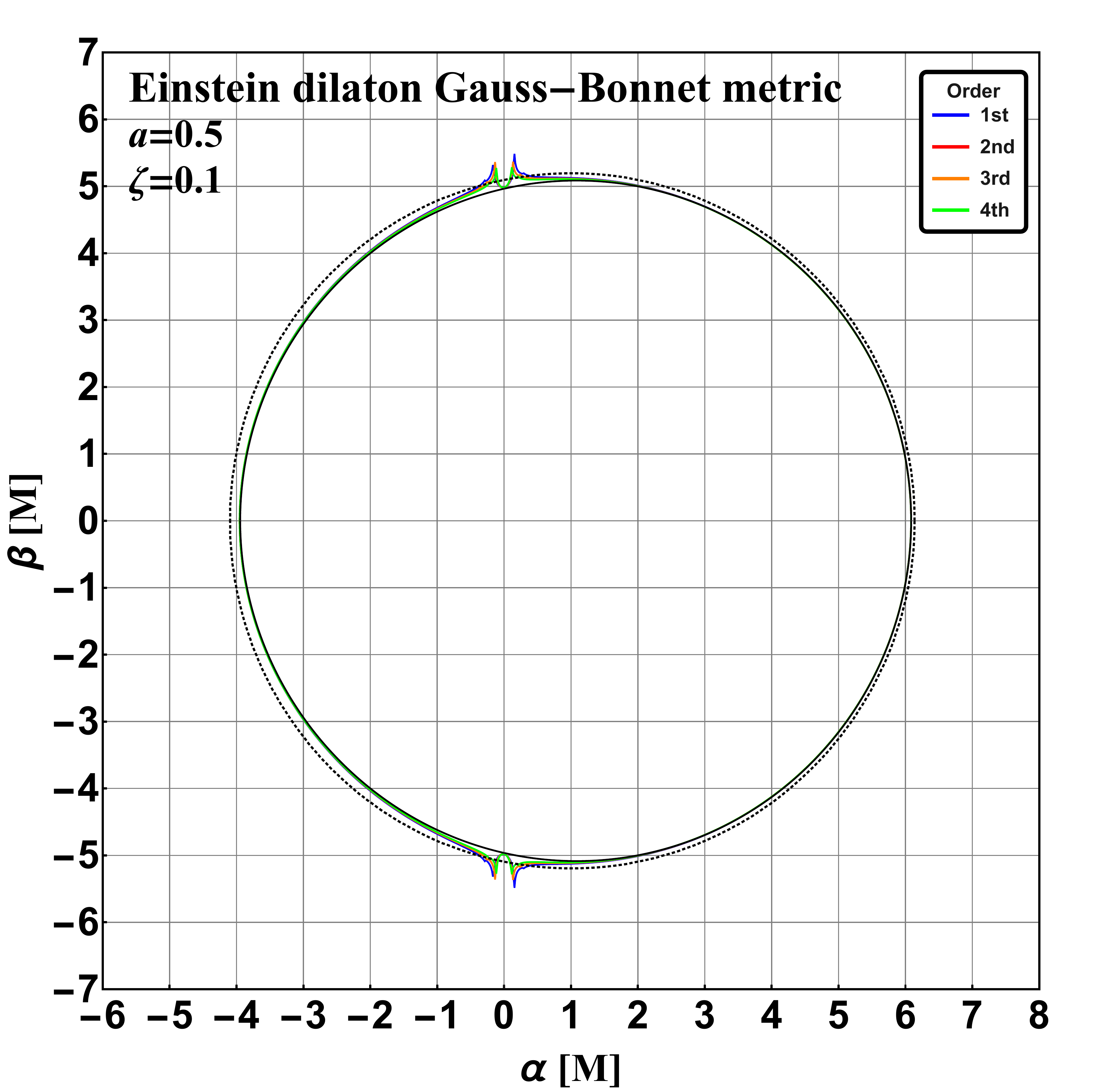}
\hskip 1.3cm
\includegraphics[width=0.40\textwidth]{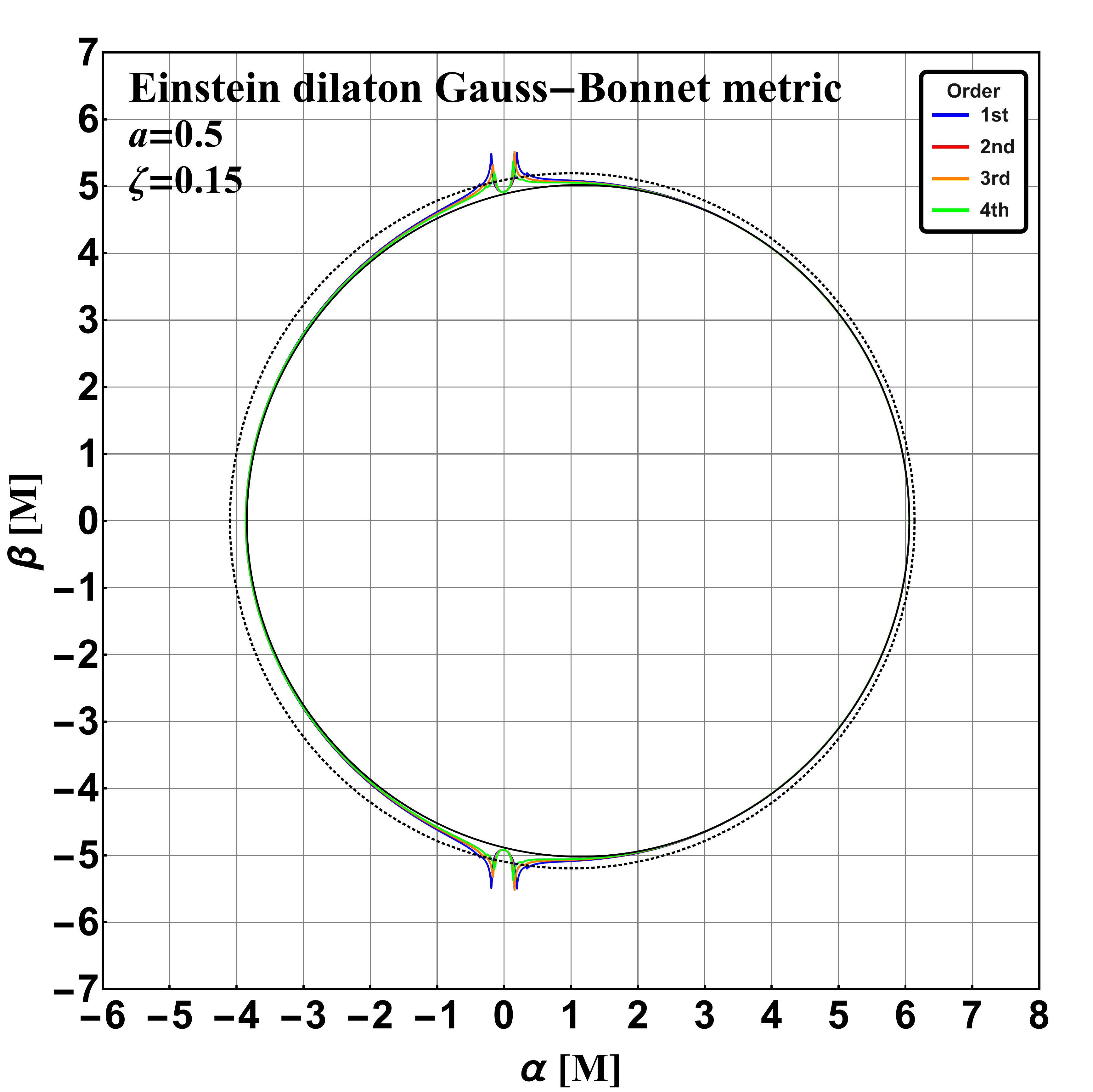} 
\vskip 0.3cm
\includegraphics[width=0.40\textwidth]{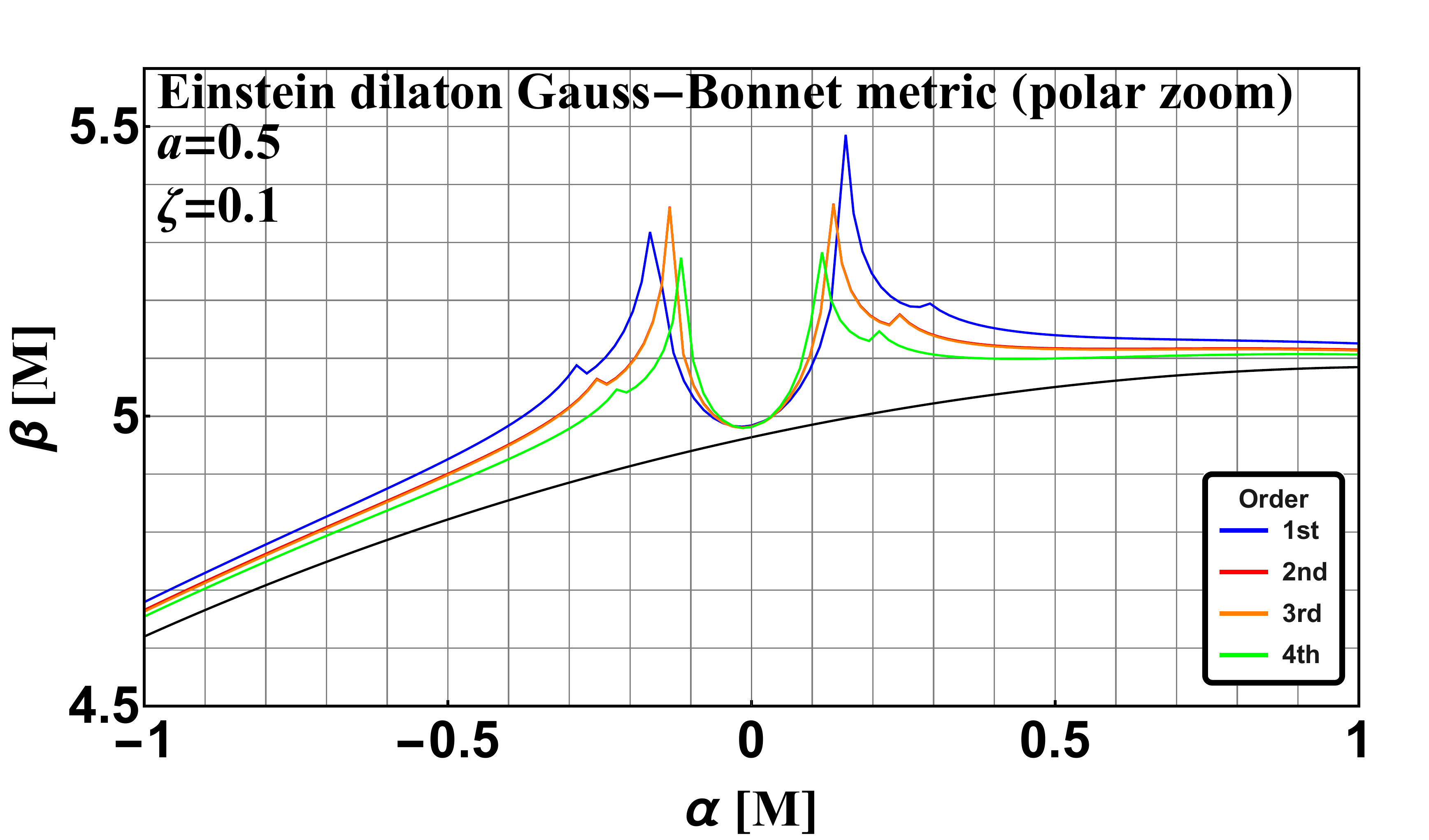}
\hskip 1.3cm
\includegraphics[width=0.40\textwidth]{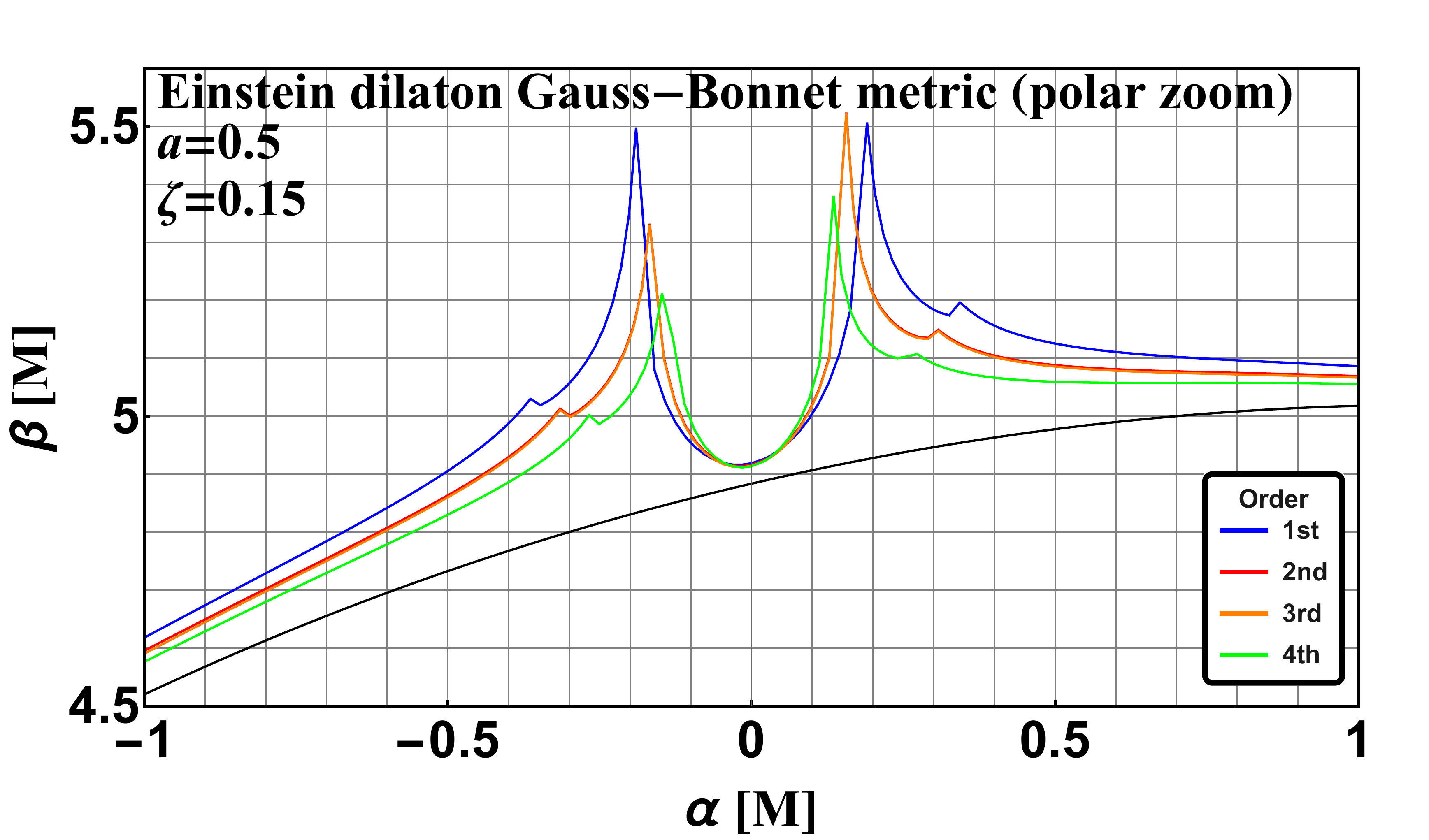} 
\vskip 0.3cm
\includegraphics[width=0.40\textwidth]{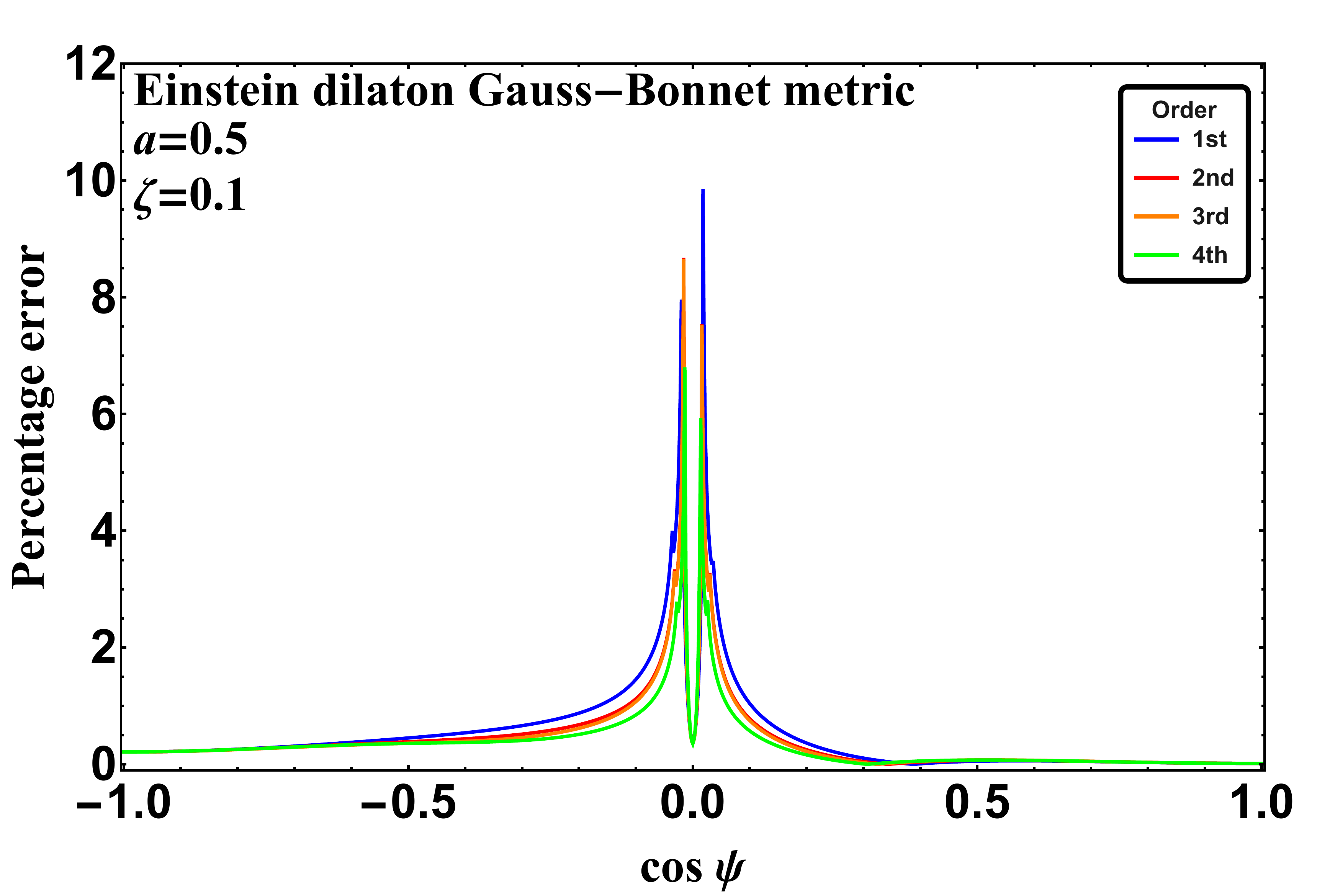}
\hskip 1.3cm
\includegraphics[width=0.40\textwidth]{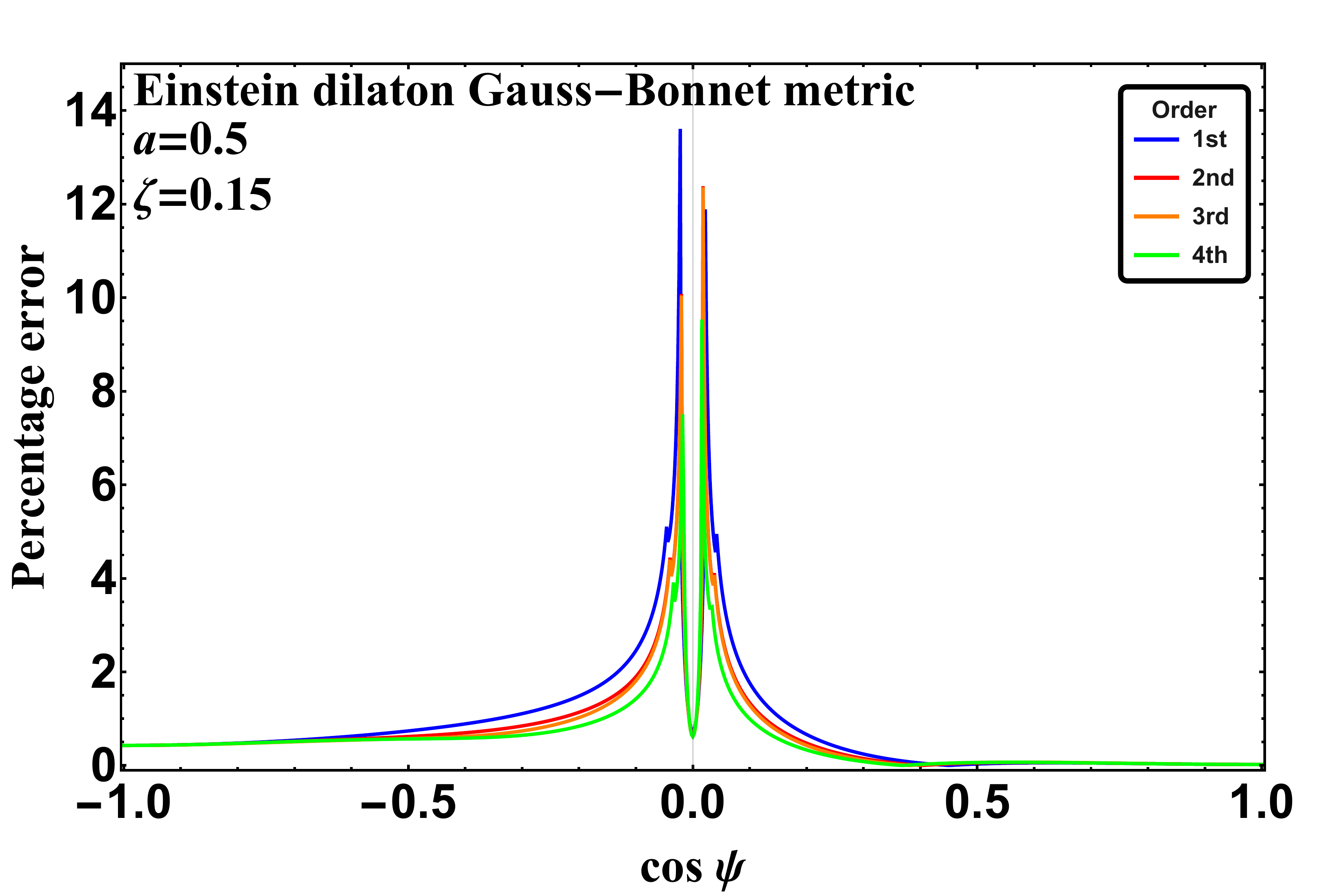}
\caption{\textit{Top row:} Shadows from the polar coordinate expansion of
  the EDGB metric with spin parameter $a=0.5$ for deformation parameters
  $\zeta=0.1$ (\textit{left panel}) and $\zeta=0.15$ (\textit{right
    panel}). \textit{Middle row:} A magnified view of the polar
  region. \textit{Bottom row:} Percentage error plots of each expansion
  order relative to the original EDGB metric. 
  Blue, red, orange and green curves denote the expansion at first, 
  second, third and fourth order in $\cos^{2}\theta$, respectively.
  For comparison, the dotted
  curve shows the shadow from a Kerr black hole with the same spin
  parameter.}
\label{fig-4}
\end{figure*}

Figure~\ref{fig-4} presents shadow calculations for the EDGB metric at
successive expansion orders in $\cos\theta$. The spin parameter is chosen
as $a=0.5$, which is near the limit of validity of the EDGB solution,
itself only derived in the literature for very small values of the spin
parameter. Two values of the deformation parameter, $\zeta$, are chosen
as $0.1$ and $0.15$. We recall that the value $\zeta=0.15$ is a critical
value, beyond which, for $a=0.5$, the EDGB metric would develop a naked
singularity. As such, these deformation parameters coupled with the
moderate spin parameter represent an extreme test of the behaviour of the
metric expansion.

As it can be seen in the top row of Fig.~\ref{fig-4}, near the equatorial
plane, and for the majority of the region away from the poles, the
expansion at all orders is in close agreement with the exact EDGB metric.
However, considering the $\alpha<0$ portion of the shadow, the shadow in the
expanded metric begins to differ more significantly from the analytic one
when approaching the polar region, before re-converging towards precisely
$\theta=\pi/2$. This behaviour is then mirrored in the $\alpha>0$ portion of
the shadow. 

To see this more clearly, a magnified view of the
neighbourhood of the polar region is presented in the middle row of
Fig.~\ref{fig-4}. The expansion exhibits some mild oscillatory behaviour
as it approaches the pole, diminishing as the order of the expansion is
increased. The red (fourth order - second order in $\cos^{2}\theta$) and orange (sixth order - third order in $\cos^{2}\theta$) curves are
almost indistinguishable in this region, but upon closer inspection of
the image, the red curve is always above the orange curve.
The small discrepancy observed in the polar region is not due to any singular behaviour in the expansion of the EDGB metric itself, but merely a reflection of the fact that the expansion is made near the equatorial plane and, in this instance, yields the largest error near the poles.

The bottom row of Fig.~\ref{fig-4} reports the percentage relative error
and, as expected, it demonstrates that the expansion worsens near the
pole and the error may be as large as $14~\%$ in the case of the second
order expansion. However, in this instance a global measure of the
shadow, namely the half-shadow area, is perhaps more representative of
the overall performance. Table~\ref{table:EDGB} shows that for
$\zeta=0.1$ the error is always less than $1~\%$. The effect of
increasing the value of $\zeta$ is to decrease the radius of the event
horizon, and therefore the photon region and by extension the calculated
area of the half-shadow. For $\zeta=0.15$ the error is $1.45~\%$. In both
cases the convergence of the expansion as the order is increased is
clear.

At this point it is important to emphasise that the EDGB metric discussed
above is essentially non-Kerr in the sense that it refers to a
non-Einsteinian theory of gravity built out of the Einstein-Hilbert
action and of a scalar field coupled to the higher curvature Gauss-Bonnet
term. Furthermore, the values of the coupling constants $\chi$ and
$\zeta$ are chosen in the test to be rather large so as to produce a
smooth but non-negligible deformation of the spacetime geometry. In this
respect, the EDGB metric represents not only a very good (and
challenging) test for the parameterisation approach, but also a demanding
benchmark for the new ray-tracing formalism.

%%%%%%%%%%%%%%%%%% FIGURE 4 %%%%%%%%%%%%%%%%%%%%
%%%%%%%%%%%%%%%% END OF FIGURE 4 %%%%%%%%%%%%%%%%%

\subsection{Johannsen-Psaltis Metric}

As a final test of the metric parameterisation we consider the
Johannsen-Psaltis metric \cite{Johannsen:2011dh}, where, for simplicity,
we take $\varepsilon_{3}$ as the only non-zero deformation parameter. In
this case, the Johannsen-Psaltis metric may be written as
\begin{eqnarray}
{d}s^{2} &=& -\frac{f-w^{2}\sin^{2}\vartheta}{\kappa^2}{d}t^{2} + \frac{\beta^2\sigma}{f}{d}\rho^{2} + \rho^{2} \sigma \ \! {d}\vartheta^{2} \nonumber \\
&& - 2w \ \! \rho \sin^{2}\vartheta \ \! {d}t \ \! {d} \ \! \phi +  \rho^{2} \kappa^2 \ \! \sin^{2}\vartheta \ \! {d}\phi^{2} \,, \label{JP_metric}
\end{eqnarray}
where
\begin{subequations}
\begin{eqnarray}
\sigma &=& 1+\frac{a^2}{\rho^2}\cos^2\vartheta\,, \\
\kappa^2 &=& \frac{(\rho^2+a^2)^2-a^2\sin^2\vartheta(\rho^2-2M \rho+a^2)}{\rho^4\sigma}\nonumber\\&&+h\frac{a^2\sin^2\vartheta(\rho^2+2M \rho+a^2\cos^2\vartheta)}{\rho^4\sigma}\,, \\
\beta &=& 1+h\,, \\
f &=& (1+h)\frac{\rho^2-2M \rho+a^2+a^2 h\sin^2\vartheta}{\rho^2} \,, \\
w &=& \frac{2 a M(1+h)}{\rho^2\sigma} \,,\\
h &=& \varepsilon_{3}\frac{M^{3}}{\rho^3\sigma^2} \,.
\end{eqnarray}
\end{subequations}

Following Ref. \cite{Konoplya:2016jvv}, we obtain the new coordinates as
a series expansion in terms of $\cos^2\vartheta$ as
\begin{eqnarray}
\cos\theta&=&\left(1+\varepsilon_{3}\frac{a^2M^{3}}{\rho^5}\right)^{-1/2}\cos\vartheta+{\cal O}(\cos^3\vartheta) \,,\\
r&=&\rho\left(1+\varepsilon_{3}\frac{a^2M^{3}}{\rho^5}\right)^{1/2}+{\cal O}(\cos^2\vartheta)\,.
\end{eqnarray}

%%%%%%%%%%%%%%%%%%%% FIGURE 5 %%%%%%%%%%%%%%%%%%%%%%%
\begin{figure*}
\begin{center}
\raisebox{0.00cm}{\includegraphics[width=0.35\textwidth]{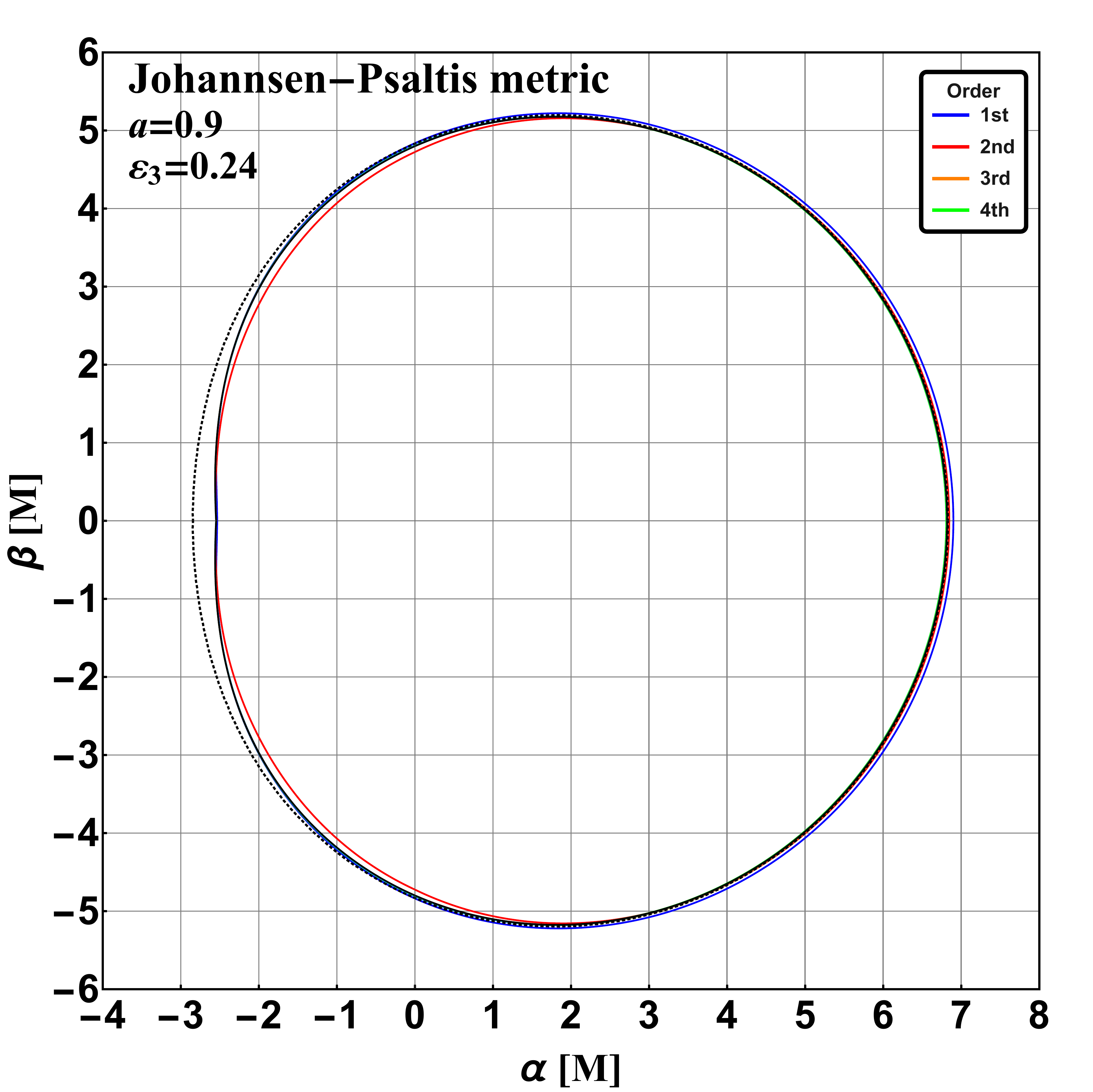}}
\hskip 1.0cm
\raisebox{0.75cm}{\includegraphics[width=0.40\textwidth]{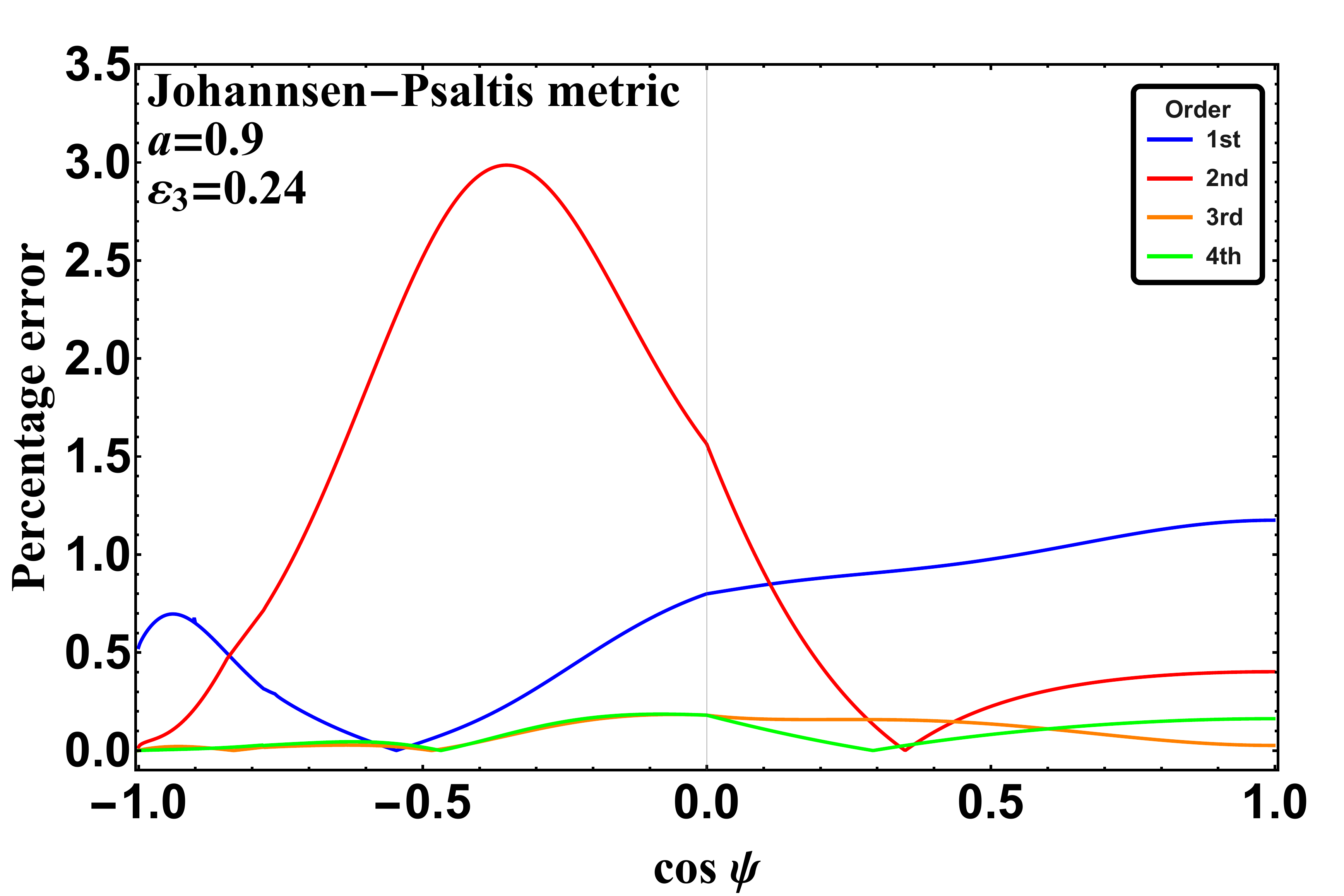}}
\vskip 0.5cm
\raisebox{0.00cm}{\includegraphics[width=0.35\textwidth]{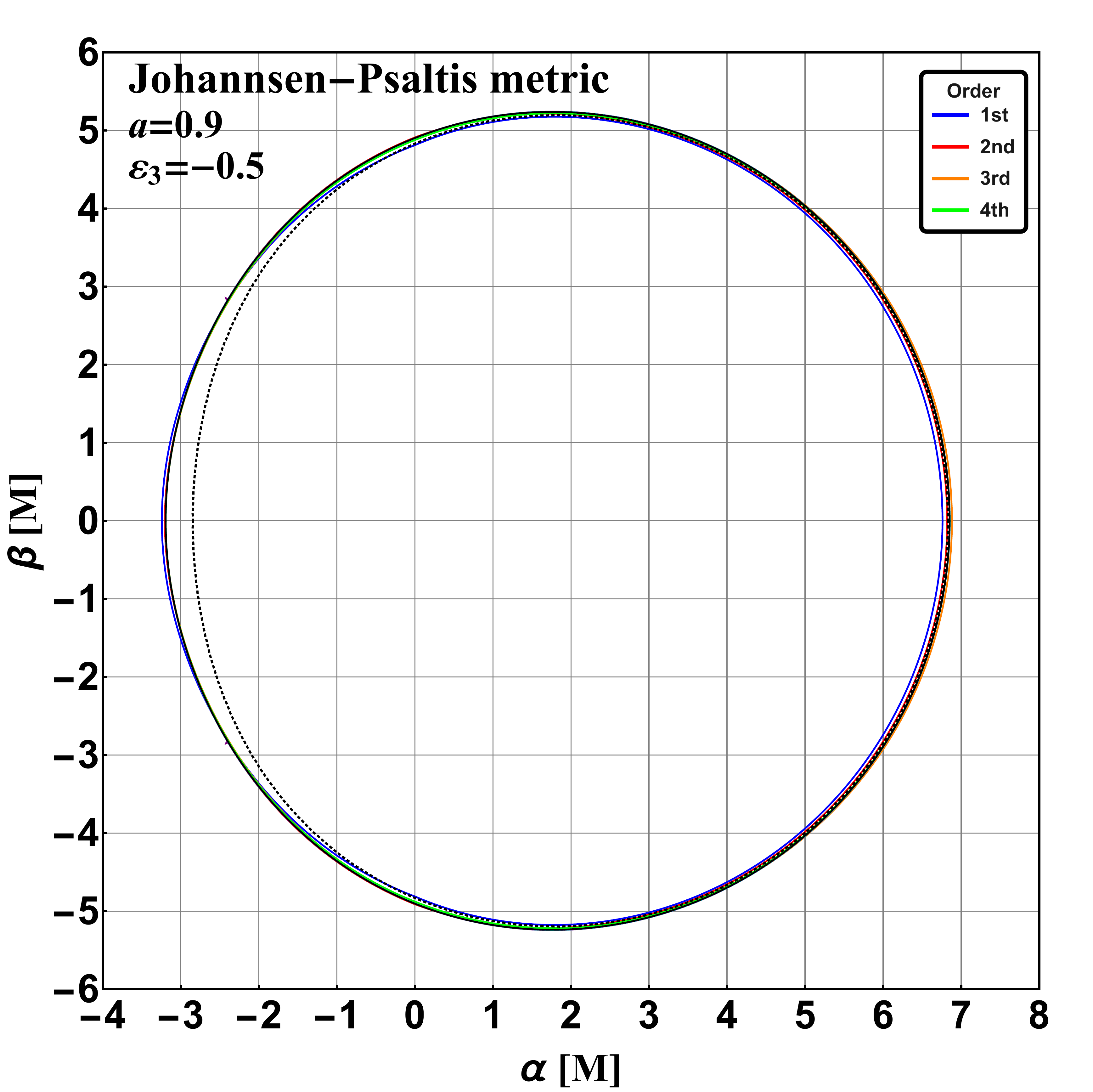}}
\hskip 1.0cm
\raisebox{0.75cm}{\includegraphics[width=0.40\textwidth]{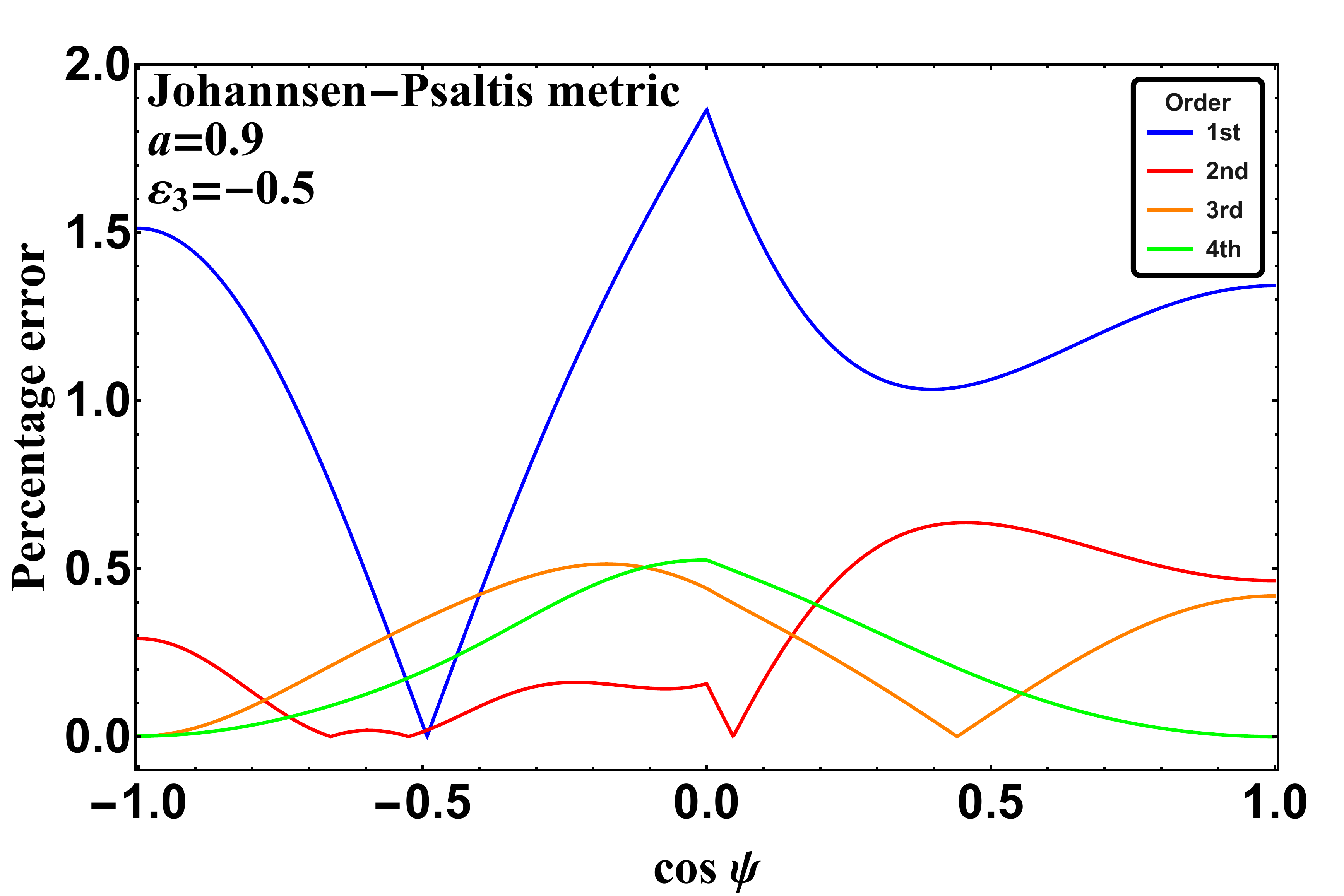}} 
\vskip 0.5cm
\raisebox{0.00cm}{\includegraphics[width=0.35\textwidth]{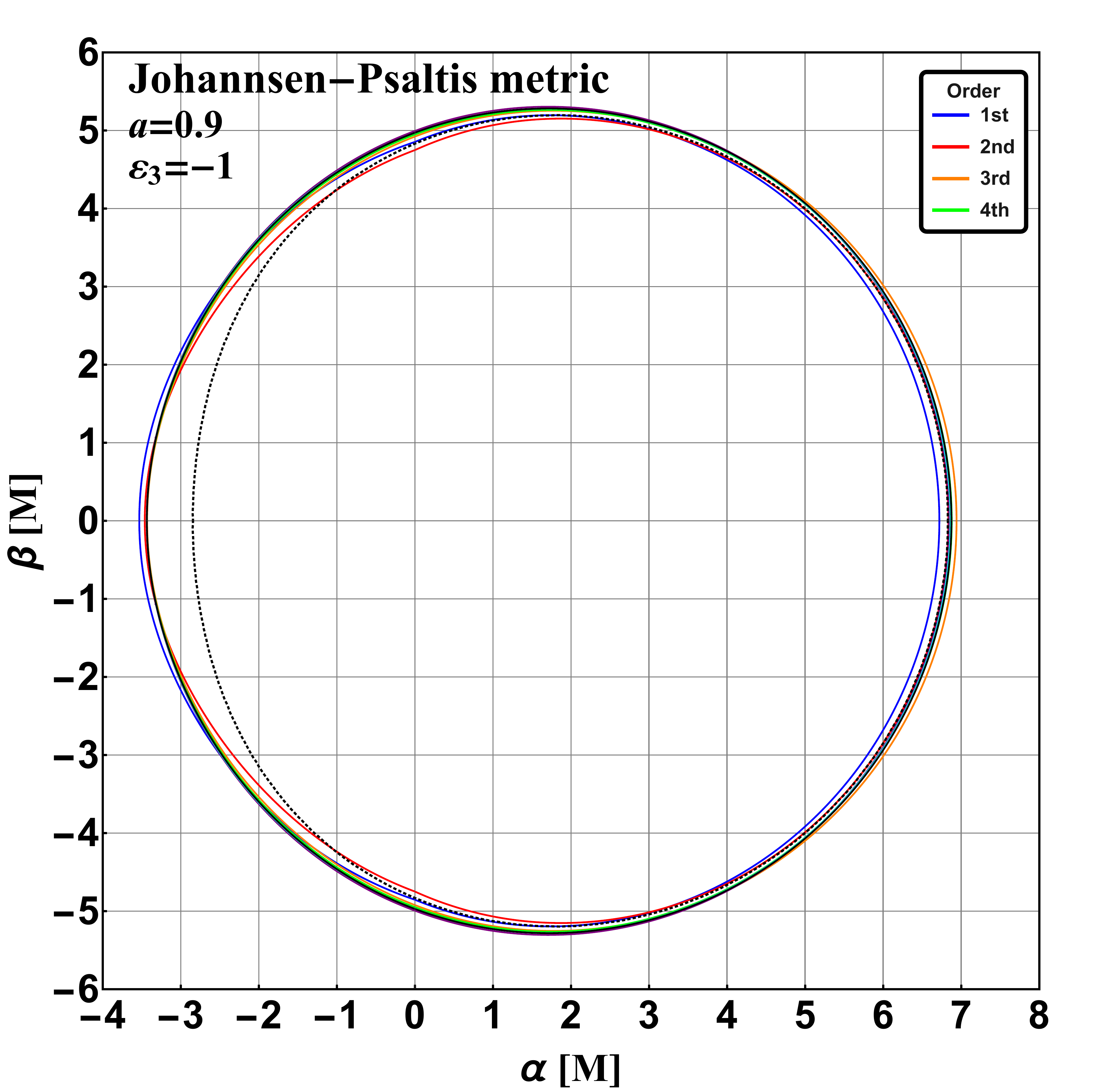}} 
\hskip 1.0cm
\raisebox{0.75cm}{\includegraphics[width=0.40\textwidth]{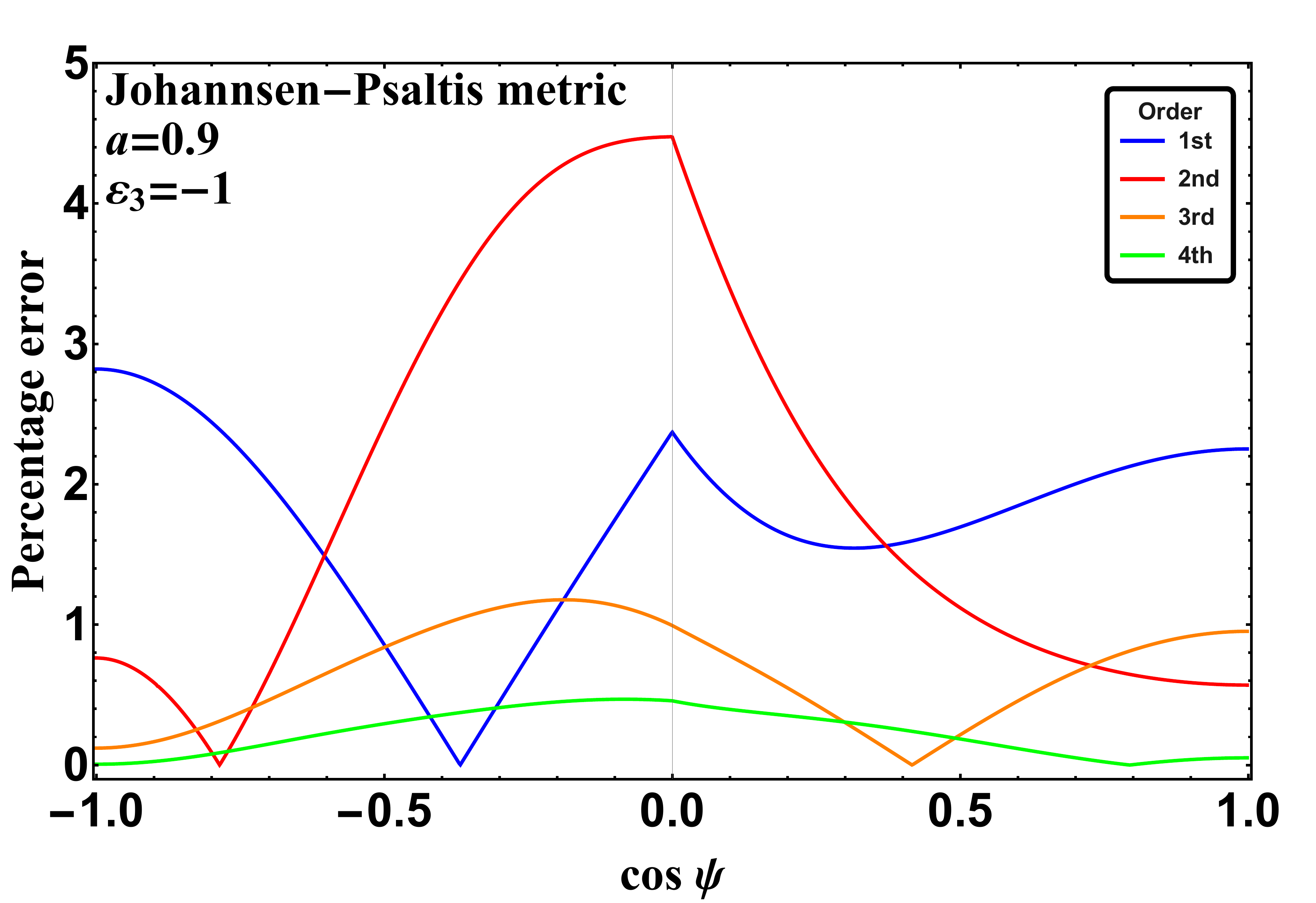}}
\end{center}
\caption{\textit{Left column:} Shadows from the radial expansion of the
  Johannsen-Psaltis metric with spin parameter $a=0.9$ for
  $\varepsilon_{3}=0.24$ (\textit{top}), $\varepsilon_{3}=-0.5$
  (\textit{middle}) and $\varepsilon_{3}=-1$ (\textit{bottom}). Only the
  first four orders of the expansion are shown. \textit{Right column:}
  Percentage errors of each expansion order relative to the original
  Johannsen-Psaltis metric. Blue, red, orange and green curves denote the expansion at first, 
  second, third and fourth order in $\cos^{2}\theta$, respectively. For comparison, the dotted curve shows the
  shadow from a Kerr black hole with the same spin parameter.}
\label{fig-5}
\end{figure*}
%%%%%%%%%%%%%%%%%% END OF FIGURE 5 %%%%%%%%%%%%%%%%%%%%

Although this series expansion cannot be inverted analytically, one may
calculate as many coefficients in the double expansion as is necessary.
More specifically, the radial expansion has been considered from the
first through to the eighth order, while the polar expansion was kept
fixed at fourth order in $\cos\theta$. Figure~\ref{fig-5} (left column)
presents shadows of the first four orders of the expanded form of the
Johannsen-Psaltis metric (red, blue, green and orange lines), along with
the shadow computed from the analytic metric, which is shown for
comparison (black line). As in the previous figures, the right column
displays the corresponding relative errors of each shadow as a function
of the polar angle $\psi$.

We have here considered a value $\varepsilon_{3}=0.24$ as this enables the
deformations in both the spin and $\varepsilon_{3}$ to be large. To increase
$\varepsilon_{3}$ further (without creating a naked singularity) would
require decreasing the spin parameter. Since the metric expansion
reproduces Kerr exactly in the equatorial plane and is increasingly
accurate for decreasing values of $a$, and since for fixed spin
parameter, increasing $\varepsilon_{3}$ has the effect of decreasing the
shadow size, the aforementioned choice of $\varepsilon_{3}=0.24$ proves
useful when scrutinising the performance of the expansion.

It is clear from the top right panel of Fig.~\ref{fig-5}, which refers to
$\varepsilon_{3}=0.24$, that the series expansion converges more slowly,
with the second order expansion error (red) proving worse than that at
first order for approximately $-0.85 \le \cos\psi \le 0.15$, \ie roughly
half of the shadow boundary. Similar behaviour is also observed for the
third order (orange) and fourth-order (green) curves. This behaviour
continues for the fifth through to eighth orders (not shown in the figure
for clarity). However, when considering the half-shadow areas in
Table~\ref{table:JP_1} the convergence is apparent, except for the $\{6,4\}$
expansion, for which we find in Table~\ref{table:JP_2} that $\epsilon_{6,4}> \epsilon_{5,4}$.
For this reason the expansion was continued up to eighth order.

Decreasing the value of $\varepsilon_{3}$ to $\varepsilon_{3}=-0.5$, the shadow
area grows and this is shown in the left panel of the middle row of
Fig.~\ref{fig-5}. Note that the first-order expansion is now the least
accurate, as one would expect (although more accurate that for
$\varepsilon_{3}=0.24$). However, inspecting the corresponding right panel
for the percentage error reveals that the second-order expansion is, in
the region roughly $-0.75\le\cos\psi\le 0.15$, more accurate than both
the third and fourth-order expansions. For $\varepsilon_{3}=-1$ this trend
continues and it is hard to discern just by looking at the first four
expansion orders whether convergence is present.

To address this issue and clarify matters,
Tables~\ref{table:JP_1}--\ref{table:JP_2} report, respectively, the half-shadow areas and their corresponding percentage errors as they are
calculated up to the eighth order in the radial coordinate and at second
order in $\cos\theta$. Whilst the shadow curves are calculated for all orders, 
they are only displayed up to fourth order in Fig.~\ref{fig-5} since they are 
visually indistinguishable from the fourth order case. 
In this way it is evident upon inspecting
Table~\ref{table:JP_2} that the expansion is indeed convergent. For all
values of $\varepsilon_{3}$ it is found that $\epsilon_{8,4} <
\epsilon_{8,2}$, and thus the expansion is convergent.

Two remarks should be made at this point. First, to appreciate the
oscillatory behaviour in the convergence of the shadow one should recall
that parameterisation is the result of a \textit{double} expansion in the
$r$ and $\theta$ directions and that the convergence in each direction
occurs at a different ``rate''. Therefore, it is perfectly possible at
some order $n$ of the expansion in one direction and order $m$ in the
other direction, one may have a situation where further extension of the
expansion in only one of these directions will lead to worse results. In
this case, this behaviour simply indicates that an increase of the
expansion order in one direction should be accompanied also by the
equivalent increase for the expansion in the other direction. Second, for
all the values of $\varepsilon_3$ considered here, the relative error
between the expanded and ``analytic'' shadows is $\lesssim 1\%$ already
at the third order and this is already much smaller than the precision at
which the measurements of the shadow will be carried out in practice.

In summary, although in principle the convergence of the parameterisation
cannot be analysed in the general case, the examples considered with the
Kerr-Sen (see Table~\ref{table:Sen}) and EDGB (see
Table~\ref{table:EDGB}) black holes show that a clear convergence between
the shadow from the expanded metric and that from the analytic
metric. Furthermore, for the Johannsen-Psaltis metric, we observe
convergence in both directions even for very large values of the
deformation parameter $\varepsilon_{3}$ (see Table~\ref{table:JP_2}). In
particular, large positive values of this parameter, combined with rapid
rotation, correspond to a shape of the event horizon that is highly
prolate and close to its extreme form (further increases of this
deformation lead to discontinuity of the horizon). At the same time, it
is possible to study smaller negative values of the deformation parameter
(which correspond to more oblate horizon shapes) and that, in the case of
rapid rotation, yield rather exotic event-horizon shapes, akin to a
dumbbell. Yet, convergence is also observed for such exotic
configurations, thus representing a convincing evidence that our
parameterisation suitably represents a wide class of axisymmetric
black hole spacetimes.

%--------------------------------------------------------------
\section{Conclusions}
\label{sec:c}
%--------------------------------------------------------------

We have introduced and subsequently employed a new method for performing
general-relativistic ray-tracing calculations in order to calculate the
black hole shadow images from a new parameterisation of any axisymmetric
black hole metric. This new parameterisation can, with a small number of
terms, represent any general stationary and axisymmetric black hole in
any metric theory of gravity. We investigated and verified the
effectiveness of this parameterisation for successive orders in the
expansion, demonstrating both its convergence and accuracy for three
different spacetimes:

\begin{enumerate}

\item The Kerr-Sen metric, fixed at second order in the polar expansion
  and varied from first through to fourth order in the radial expansion
  (the Kerr black hole is exactly reproduced at second order in the polar
  direction).

\item A regular form of the EDGB metric, itself obtained from an
  expansion in terms of the parameters $\chi$ and $\zeta$. This regular
  solution is approximate but converges, to any desired accuracy, to the
  original approximate EDGB solution that diverges at $r=2M$. The
  expansion was purely polar and varied from first through to fourth
  order (in $\cos^{2}\theta$) in the polar direction.

\item The Johannsen-Psaltis metric, represented as a double expansion in
  both the polar and radial directions. The expansion was fixed at second
  order in the polar direction and varied from first to eighth order in
  the radial direction.

\end{enumerate}

For all the aforementioned metrics, we chose values of the spin parameter
and metric deformation parameters to be as extremal as possible whilst
still ensuring the existence of an event horizon (\ie avoiding the
appearance of a naked singularity). We performed three tests for each
expansion order of each metric, calculating: (i) the black hole shadow
polar curve obtained at each order, (ii) the error relative to the exact
metric along the shadow curve, and (iii) the error of the half-area of
the black hole shadow with respect to that obtained from the exact
metric.

Test (i) provided a qualitative comparison of the performance of the
expansion as the order was increased, while test (ii) provided a
quantification of the performance of the parameterisation everywhere
along the shadow boundary. This test, in particular, provided an
understanding of how well the parameterisation represents the spacetime,
for example, in the equatorial plane and at the poles. Finally, test
(iii) verified the excellent convergence behaviour and accuracy of the
parameterisation as the expansion order was increased. We demonstrated
that by increasing the order of polar and radial expansions the
spacetime under consideration can be represented to essentially any
desired accuracy\footnote{\texttt{Mathematica} notebooks containing the
  relevant coefficients for the metrics considered in this study may be
  found in the supplementary material accompanying this paper.}.

Accurate calculations of black hole shadows in parameterised metrics
represent a stringent test of parameterised representations of metric
theories of gravity. Photons which delineate the shadow boundary pass
very close to the event horizon and are subject to the steepest gradients
of the gravitational potentials. Hence, accurately reproducing the
behaviour of the spacetime in these regions lends credence to the
prospect of employing this parameterisation framework to investigate not
only black hole solutions in other metric theories of gravity, but to
also perform the detailed radiative transport calculations required to
investigate physical processes in other theories of gravity. Such
calculations will prove useful for the interpretation of upcoming sub-mm
VLBI observations from Sgr A* and for testing the Kerr black hole
hypothesis.

%%%%%%%%%%%%%%%%%% TABLE 1 %%%%%%%%%%%%%%%%%%%%%
\begin{table*}
\setlength{\tabcolsep}{7pt}
\vspace*{-1mm}
\caption{Table of the half-shadow area, $A_{n,m}$ (in units of $M^{2}$),
  and its corresponding percentage error (with respect to the analytic
  Kerr-Sen metric), $\epsilon_{n,m}$, of each expansion order
  $\{n,m\}$ of the Kerr-Sen metric. Indices $n$ and $m$ correspond,
  respectively, to the order of the \textit{radial} and $\cos\theta$
  expansions. The exact area obtained from the analytic metric is denoted
  by $A_{\mathrm{exact}}$. Numbers within square brackets denote
  multiplicative powers of 10. For the Kerr-Sen metric $\cos\theta$ is
  fixed at second order and the \textit{radial} expansion is varied up to
  fourth order.}
\centering
\vspace{1mm}
\begin{tabular}{| c | c | c | l l l l | l l l l |}             % centered columns (11 columns)
\hline                         %inserts horizontal line
$a$ & $b$ & ${A_{\mathrm{\mathbf{exact}}}}$ & ${A_{1,2}}$ &
${A_{2,2}}$ & ${A_{3,2}}$ & ${A_{4,2}}$ & ${\epsilon_{1,2}}$ &
${\epsilon_{2,2}}$ & ${\epsilon_{3,2}}$ & ${\epsilon_{4,2}}$
\\ 
\hline 
\multirow{2}*{$0.2$} & $0.5$ & $73.5640$ & $73.5641$ & $73.5642$  & $73.5640$  & $73.5640$  & $1.06[-4]$  & $3.22[-4]$ & $3.08[-5]$ & $4.85[-6]$ \\[0.0ex]
\cline{2-11}                 
 & $1.0$ & $110.576$ & $110.601$ & $110.581$ & $110.575$ & $110.576$ & $2.32[-2]$ & $4.38[-3]$ & $4.29[-4]$ & $2.75[-6]$ \\ [0.0ex]
\cline{1-11}
\multirow{2}*{$0.5$} & $0.5$ & $72.6472$ & $72.6289$ & $72.6478$  & $72.6471$  & $72.6472$  & $2.52[-2]$  & $8.20[-4]$ & $1.29[-4]$ & $8.18[-6]$ \\[0.0ex]
\cline{2-11}
 & $1.0$ & $109.237$ & $109.241$ & $109.240$ & $109.235$ & $109.237$ & $3.83[-3]$ & $2.91[-3]$ & $1.28[-3]$ & $6.11[-5]$ \\ [0.0ex]
\cline{1-11}
\multirow{2}*{$0.95$} & $0.5$ & $68.3328$ & $67.9290$ & $68.4379$  & $68.3317$  & $68.3328$  & $5.91[-1]$  & $1.54[-1]$ & $1.49[-3]$ & $1.10[-4]$ \\[0.0ex]
\cline{2-11}
 & $1.0$ & $102.953$ & $102.512$ & $103.081$ & $102.940$ & $102.954$ & $4.28[-1]$ & $1.25[-1]$ & $1.25[-2]$ & $9.51[-4]$ \\ [0.0ex]
 \cline{1-11}
\multirow{2}*{$0.998$} & $0.5$ & $66.7159$ & $65.8826$ & $67.1017$  & $66.7136$  & $66.7160$  & $1.25[+0]$  & $5.78[-1]$ & $3.41[-3]$ & $1.90[-4]$ \\[0.0ex]
\cline{2-11}
 & $1.0$ & $100.620$ & $99.7024$ & $101.032$ & $100.588$ & $100.622$ & $9.12[-1]$ & $4.10[-1]$ & $3.13[-2]$ & $1.77[-3]$ \\[0.0ex]
\hline                       
\end{tabular}
\label{table:Sen}            
\end{table*}
%%%%%%%%%%%%%%%%%% END OF TABLE 1 %%%%%%%%%%%%%%%%%%%%

%%%%%%%%%%%%%%%% TABLE 2 %%%%%%%%%%%%%%%%%
\begin{table*}
\setlength{\tabcolsep}{7pt}
\vspace*{0mm}
\caption{Table of coefficients for the non-divergent EDGB metric.}      % title of Table
\centering 
\vspace{1mm}
\begin{tabular}{| c | c | c | c c c | c c c |}               % centered columns (9 columns)
\hline                         
${k}$ & ${c_{k}}$ & ${w_{k}}$ & ${f_{k,1}}$ & ${f_{k,2}}$ & ${f_{k,3}}$ & ${\beta_{k,1}}$ & ${\beta_{k,2}}$ & ${\beta_{k,3}}$  \\ 
\hline
$0$ & $-{4463}/{875}$ & $-9$ & ${3019}/{875}$ & $-{3048}/{875}$  & ${26}/{3}$  & $2$  & $5$  & $-14$ \\
\cline{1-9}
$1$ & $-{2074}/{175}$ & $-140$ & ${11201}/{1750}$ & $-{18551}/{5250}$ & ${22}/{5}$ & $11$ & ${139}/{15}$ & $-{128}/{5}$ \\ [0.0ex]
\cline{1-9}
$2$ & $-{266911}/{12250}$ & $-90$ & $-{1497089}/{36750}$ & ${838039}/{110250}$ & ${32}/{5}$ & ${2767}/{15}$ & $-{907}/{45}$ & $-48$  \\[0.0ex]
\cline{1-9}
$3$ & $-{12673}/{525}$ & $-144$ & ${30316}/{3675}$ & $-{253756}/{11025}$ & $-{80}/{3}$ & $-{208}/{5}$ & ${616}/{5}$ & 0 \\[0.0ex]
\cline{1-9}
$4$ & ${12371}/{245}$ & $400$ & $-{26233}/{245}$ & ${1917}/{245}$ & 0 & ${1658}/{5}$ & ${2102}/{15}$ & $0$  \\[0.0ex]
\cline{1-9}
$5$ & ${3254}/{35}$ & ---  & ${9214}/{21}$ & $-{20422}/{315}$ & $0$ & $-{26288}/{15}$ & ${28688}/{45}$ & $0$ \\[0.0ex]
\cline{1-2}\cline{3-9}
$6$ & ${2536}/{15}$ & ---  & $-{6136}/{15}$ & ${7336}/{45}$ & $0$ & $2160$ & $-720$ & $0$ \\[0.0ex]
\cline{1-2}\cline{3-9}
$7$ & $-240$ & --- & $240$ & $-80$ & $0$ & --- & --- & --- \\[0.0ex]
\hline                                 
\end{tabular}
\label{table:coefficients}                   
\end{table*}
%%%%%%%%%%%%%% END OF TABLE 2 %%%%%%%%%%%%%%%

%%%%%%%%%%%%%%%%%%%% TABLE 3 %%%%%%%%%%%%%%%%%%%%%%%
\begin{table*}
\setlength{\tabcolsep}{7pt}
\vspace*{-2mm}
\caption{Table of half-shadow areas and their corresponding percentage
  errors for the first four successive expansion orders of the EDGB
  metric.  The expansion considered in this case is in terms of
  $\cos^{2}\theta$ only.}  \centering
\vspace{1mm}
\begin{tabular}{| c | c | c | l l l l | l l l l |}             
\hline                         
$a$ & $\zeta$ & ${A_{\mathrm{exact}}}$ & ${A_{0,2}}$ & ${A_{0,4}}$ & ${A_{0,6}}$ & ${A_{0,8}}$ & ${\epsilon_{0,2}}$ & ${\epsilon_{0,4}}$ & ${\epsilon_{0,6}}$ & ${\epsilon_{0,8}}$  \\ 
\hline
\multirow{2}*{$0.5$} & $0.1$ & $39.9885$ & $40.3445$ & $40.2700$  & $40.2559$  & $40.2088$  & $8.90[-1]$  & $7.04[-1]$ & $6.69[-1]$ & $5.51[-1]$ \\
\cline{2-11}
 & $0.15$ & $38.9264$ & $39.4915$ & $39.3773$ & $39.3536$ & $39.2825$ & $1.45[+0]$ & $1.16[+0]$ & $1.10[+0]$ & $9.15[-1]$ \\ 
\hline
\end{tabular}
\label{table:EDGB}
\end{table*}
\begin{table*}
\setlength{\tabcolsep}{5pt}
\vspace*{-2mm}
\centering
\caption{Table of half-shadow areas for successive expansion orders of the Johannsen-Psaltis metric.
The expansion is fixed at fourth order for $\cos\theta$ whilst the \textit{radial} expansion is considered up to eighth order.
Note that for the eighth order radial expansion the value of $A_{8,2}$ (\ie second order in $\cos\theta$) is included to compare with $A_{8,4}$.
}
\centering
\vspace{1mm}
\begin{tabular}{| c | c | c | l l l l l l l l l |}
\hline          \vspace*{-1mm}             
$a$ & $\varepsilon_{3}$ & ${A_{\mathrm{exact}}}$ & ${A_{1,4}}$ & ${A_{2,4}}$ & ${A_{3,4}}$ & ${A_{4,4}}$ & ${A_{5,4}}$ & ${A_{6,4}}$ & ${A_{7,4}}$ & ${A_{8,2}}$ & ${A_{8,4}}$  \\ 
  &  &  &  &  &  &  &  &  &  & &  \\ [-1ex] 
 \cline{4-12}              
\hline
\\ [-3.15ex]
\multirow{3}*{$0.9$} & \multirow{1}*{$0.24$} & \multirow{1}*{$38.8979$} & $39.5424$ & $38.5721$  & $38.9766$  & $38.8723$  & $38.9158$  & $38.9175$ & $38.8838$ & $38.9171$ & $38.8888$ \\[0.0ex]                 
\cline{4-12} \\ [-3.15ex]
\cline{2-12} \\ [-3.15ex]
     & \multirow{1}*{$-0.5$} & \multirow{1}*{$41.2918$} & $40.4947$ & $41.0235$  & $41.2720$  & $41.1276$  & $41.3215$  & $41.3076$ & $41.2778$ & $41.2747$ & $41.2880$ \\[0.0ex]                 
\cline{2-12} \\ [-3.15ex]
     & \multirow{1}*{$-1.0$} & \multirow{1}*{$42.5892$} & $41.5207$ & $41.1449$  & $42.5444$  & $42.4225$  & $42.6546$  & $42.5539$ & $42.5356$ & $42.5678$ & $42.5806$ \\[0.0ex]                 
\hline
\end{tabular}
\label{table:JP_1}
\end{table*}
%%%%%%%%%%%%%%%%%%% END OF TABLE 4 %%%%%%%%%%%%%%%%%%%%

%%%%%%%%%%%%%%%%%%%%% TABLE 5 %%%%%%%%%%%%%%%%%%%%%%%
\begin{table*}
\setlength{\tabcolsep}{5pt}
\vspace*{-2mm}
\centering
\caption{Table of percentage errors corresponding to the half-shadow areas for successive expansion orders of the Johannsen-Psaltis metric as reported in Table \ref{table:JP_1}.
As expected, $\epsilon_{8,4} < \epsilon_{8,2}$ for all values of $\varepsilon_{3}$.
}
\centering
\vspace{1mm}
\begin{tabular}{| c | c | c l l l l l l l l |}
\hline          \vspace*{-1mm}             
$a$ & $\varepsilon_{3}$ & ${\epsilon_{1,4}}$ & ${\epsilon_{2,4}}$ & ${\epsilon_{3,4}}$ & ${\epsilon_{4,4}}$ & ${\epsilon_{5,4}}$ & ${\epsilon_{6,4}}$ & ${\epsilon_{7,4}}$ & ${\epsilon_{8,2}}$ & ${\epsilon_{8,4}}$  \\ [1ex] 
\hline
\\ [-3.15ex]
\multirow{3}*{$0.9$} 
& \multirow{1}*{$0.24$} & $1.66[+0]$ & $8.38[-1]$ & $2.02[-1]$ & $6.59[-2]$ & $4.58[-2]$ & $5.02[-2]$ & $3.63[-2]$ & $4.91[-2]$ & $2.34[-2]$ \\ [0.0ex]
& \multirow{1}*{$-0.5$} & $1.93[+0]$ & $6.50[-1]$ & $4.78[-2]$ & $3.98[-1]$ & $7.19[-2]$ & $3.82[-2]$ & $3.39[-2]$ & $4.13[-2]$ & $9.20[-3]$ \\ [0.0ex]
& \multirow{1}*{$-1.0$} & $2.51[+0]$ & $3.39[+0]$ & $1.05[-1]$ & $3.91[-1]$ & $1.54[-1]$ & $8.29[-2]$ & $1.26[-1]$ & $5.02[-2]$ & $2.00[-2]$ \\ [0.0ex]
\hline
\end{tabular}
\label{table:JP_2}
\end{table*}

\begin{acknowledgments}
It is a pleasure to thank Arne Grenzebach, Oliver Porth, Bruno Mundim,
Mariafelicia de Laurentis, Hector Olivares and Christian Fromm for
numerous discussions and comments.
We thank the anonymous referee for useful comments which helped improve the manuscript.
This work was supported by the ERC
Synergy Grant ``BlackHoleCam -- Imaging the Event Horizon of Black
Holes'' (Grant 610058). Z.Y. acknowledges support from an Alexander von
Humboldt Fellowship. A.~Z.~was partially supported by Conselho Nacional
de Desenvolvimento Cient\'ifico e Tecnol\'ogico (CNPq). This research has
made use of NASA's Astrophysics Data System.

\end{acknowledgments}

\end{document}